\documentclass[letterpaper, 10 pt, conference]{ieeeconf}%
\usepackage{algorithm2e}
\usepackage{algpseudocode}
\usepackage{caption}
\usepackage{slashbox}
\usepackage{subfig}
\usepackage{graphics}
\usepackage{epsfig}
\usepackage{mathptmx}
\usepackage{times}
\usepackage{amsmath}
\usepackage{amssymb}
\usepackage{amsfonts}
\usepackage{graphicx}
\usepackage[usenames, dvipsnames]{color}%
\providecommand{\U}[1]{\protect\rule{.1in}{.1in}}
\IEEEoverridecommandlockouts
\renewcommand{\baselinestretch}{1}

\begin{document}

\title{{\LARGE \textbf{Optimization of Ride Sharing Systems Using Event-driven
Receding Horizon Control$^{\star}$}}}
\author{Rui Chen and Christos G. Cassandras \thanks{ $^{\star}$Supported in part by
NSF under grants ECCS-1509084, CNS-1645681, and IIP-1430145, by AFOSR under
grant FA9550-15-1-0471, by the DOE under grant DE-AR0000796, by the MathWorks
and by Bosch.} \thanks{The authors are with the Division of Systems
Engineering and Center for Information and Systems Engineering, Boston
University, Brookline, MA 02446, USA \texttt{{\small \{ruic,cgc\}@bu.edu}}} }
\maketitle

\begin{abstract}
We develop an event-driven Receding Horizon
Control (RHC) scheme for a Ride Sharing System (RSS) in a transportation
network where vehicles are shared to pick up and drop off passengers so as to
minimize a weighted sum of passenger waiting and traveling times. The RSS is
modeled as a discrete event system and the event-driven nature of the
controller significantly reduces the complexity of the vehicle assignment
problem, thus enabling its real-time implementation. Simulation results using
actual city maps and real taxi traffic data illustrate the effectiveness of
the RH controller in terms of real-time implementation and performance relative
to known greedy heuristics.

\end{abstract}

\thispagestyle{empty} \pagestyle{empty}


\section{Introduction}

It has been abundantly documented that the state of transportation systems
worldwide is at a critical level. Based on the $2011$ Urban Mobility Report, the
cost of commuter delays has risen by $260$\% over the past $25$ years and $28$\% of
U.S. primary energy is now used in transportation \cite{schrank2011urban}.
Traffic congestion also leads to an increase in vehicle emissions; in large cities, as
much as $90$\% of CO emissions are due to mobile sources. Disruptive
technologies that aim at dramatically altering the transportation landscape
include vehicle connectivity and automation as well as shared personalized
transportation through emerging mobility-on-demand systems. Focusing on the
latter, the main idea of a Ride Sharing System (RSS) is to assign vehicles in
a given fleet so as to serve multiple passengers, thus effectively reducing
the total number of vehicles on a road network, hence also congestion, energy
consumption, and adverse environmental effects. 

The main objectives of a RSS
are to minimize the total Vehicle-Miles-Traveled (VMT) over a given time period
(equivalently, minimize total travel costs), to minimize the average waiting
and traveling times experienced by passengers, and to maximize the number of
satisfied RSS participants (both drivers and passengers)
\cite{agatz2012optimization}. When efficiently managed, a RSS has the
potential to reduce the total number of private vehicles in a transportation
network, hence also decreasing overall energy consumption and traffic
congestion, especially during peak hours of a day. From a passenger
standpoint, a RSS is able to offer door-to-door transportation with minimal
delays which makes traveling more convenient. From an operator's standpoint a
RSS provides a considerable revenue stream. A RSS also provides an alternative
to public transportation or can work in conjunction with it to reduce possible
low uitization of vehicles and long passenger delays.

In this paper, we concentrate on designing dynamic vehicle assignment
strategies in a RSS aiming to minimize the system-wide waiting and traveling
times of passengers. The main challenge in obtaining optimal vehicle
assignments is the complexity of the optimization problem involved in
conjunction with uncertainties such as random passenger service request times,
origins, and destinations, as well as unpredictable traffic conditions which
determine the times to pick up and drop off passengers. Algorithms used in RSS
are limited by the NP-complete nature of the underlying traveling salesman
problem \cite{chen2017hierarchical} which is a special case of the much more
complex problems encountered in RSS optimization. Therefore, a global optimal
solution for such problems is generally intractable, even in the absence of
the aforementioned uncertainties. Moreover, a critical requirement in such
algorithms is a guarantee that they can be implemented in a real-time context.

Several methods have been proposed to solve the RSS problem addressing the
waiting and traveling times of passengers. In \cite{agatz2011dynamic}, a
greedy approach is used to match vehicles to passenger requests which can on
one hand guarantee real-time assignments but, on the other, lacks performance
guarantees. The optimization algorithm in \cite{santi2014quantifying} improves
the average traveling time performance but limits the seat capacity of each
vehicle to $2$ (otherwise, the problem becomes intractable for $4$ or more
seats) and allows no dynamic allocation of new passengers after a solution is
determined. Although vehicles can be dynamically allocated to passengers in
\cite{berbeglia2010dynamic}, all pickup and drop-off events are constrained to
take place within a specified time window. The RTV-graph algorithm
\cite{alonso2017demand} can also dynamically allocate passengers, but its
complexity increases dramatically with the number of agents (passengers and
vehicles) and the seat capacity of vehicles. To address the issue of
increasing complexity with the size of a RSS, a hierarchical approach is
proposed in \cite{chen2017hierarchical} such that the system is decomposed
into smaller regions. Within a region, a mixed-integer linear programs is
formulated so as to obtain an optimal vehicle assignment over a sequence of
fixed time horizons. Although this method addresses the complexity issue, it
involves a large number of unnecessary calculations since there is no need to
always re-evaluate an optimal solution over every such horizon. Another
approach to reducing complexity, is to abstract a RSS model through passenger
and vehicle flows as in \cite{calafiore2017flow},\cite{tsao2018stochastic} and
\cite{salazar2018interaction}.
In \cite{salazar2018interaction}, for example, the interaction between
autonomous mobility-on-demand and public transportation systems is considered
so as to maximize the overall social welfare.

In order to deal with the well-known \textquotedblleft curse of
dimensionality\textquotedblright\ \cite{bertsekas2005dynamic} that
characterizes optimization problem formulations for a RSS, we adopt an
\emph{event-driven} \emph{Receding Horizon Control} (RHC) approach. This is in
the same spirit as Model Predictive Control (MPC) techniques
\cite{camacho2013model} with the added feature of exploiting the event-driven
nature of the control process in which the RHC algorithm is invoked only when
certain events occur. Therefore, compared with conventional time-driven MPC
this approach can avoid unnecessary calculations and can significantly improve
the efficiency of the RH controller by reacting to random events as they occur
in real time. The basic idea of event-driven RHC introduced in
\cite{li2006cooperative} and extended in \cite{khazaeni2016event} is to solve
an optimization problem over a given \emph{planning horizon} when an event is
observed in a way which allows vehicles to cooperate; the resulting control is
then executed over a generally shorter \emph{action horizon} defined by the
occurrence of the next event of interest to the controller. Compared to
methods such as \cite{santi2014quantifying}-\cite{alonso2017demand}, the RHC
scheme is not constrained by vehicle seating capacities and is specifically
designed to dynamically re-allocate passengers to vehicles at any time.
Moreover, compared to the time-driven strategy in \cite{chen2017hierarchical},
the event-driven RHC scheme refrains from unnecessary calculations when no
event in the RSS occurs. Finally, in contrast to models used in
\cite{tsao2018stochastic} and \cite{salazar2018interaction}, we maintain
control of every vehicle and passenger in a RSS at a microscopic level while
ensuring that real-time optimal (over each receding horizon) vehicle
assignments can be made.

The paper is organized as follows. We first present in Section II a discrete
event system model of a RSS and formulate an optimization problem aimed at
minimizing a weighted sum of passenger waiting and traveling times. Section
III first reviews the basic RHC scheme previously used and then identifies how
it is limited in the context of a RSS. This motivates the new RHC approach
described in Section IV, specifically designed for a RSS. Extensive simulation
results are given in Section V for actual maps in Ann Arbor, MI and New York
City, where, in the latter case, real taxi traffic data are used to drive the
simulation model. We conclude the paper in Section VI. \newline

\bigskip

\section{Problem Formulation}

We consider a Ride Sharing System (RSS) in a traffic network consisting of $N$
nodes $\mathcal{N}=\{1,...,N\}$ where each node corresponds to an
intersection. Nodes are connected by arcs (i.e., road segments). Thus, we view
the traffic network as a directed graph $\mathbb{G}$ which is embedded in a
two-dimensional Euclidean space and includes all points contained in every
arc, i.e., $\mathbb{G}\subset\mathbb{R}^{2}$. In this model, a node
$n\in\mathcal{N}$ is associated with a point $\nu_{n}\in\mathbb{G}$, the
actual location of this intersection in the underlying two-dimensional space.
The set of vehicles present in the RSS at time $t$ is $\mathcal{A}(t)$, where
the index $j\in\mathcal{A}(t)$ will be used to uniquely denote a vehicle, and
let $A(t)=|\mathcal{A}(t)|$. The set of passengers is $\mathcal{P}(t)$, where
the index $i$ will be used to uniquely denote a passenger, and let
$P(t)=|\mathcal{P}(t)|$. Note that $\mathcal{A}(t)$ is time-varying since
vehicles may enter or leave the RSS at any time and the same is true for
$\mathcal{P}(t)$.

There are two points in $\mathbb{G}$ associated with each passenger $i$,
denoted by $o_{i},r_{i}\in\mathbb{G}$: $o_{i}$ is the origin where the
passenger issues a service request (pickup point) and $r_{i}$ is the
passenger's destination (drop-off point). Let ${O}(t)=\{o_{1},...,o_{P}\}$ be
the set of all passenger origins and ${R}(t)=\{r_{1},...,r_{P}\}$ the
corresponding destination set. Vehicles pick up passengers and deliver them to
their destinations according to some policy. We assume that the times when
vehicles join the RSS are not known in advance, but they become known as a
vehicle joins the system. Similarly, the times when passenger service requests
occur are random and their destinations become known only upon being picked up.

\begin{figure}[pt]
\centering
\includegraphics[scale=0.3]{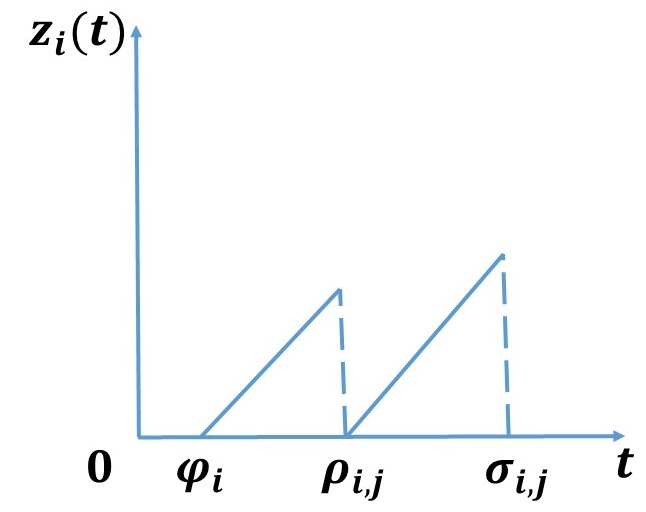}\caption{A typical sample path of
passenger $i$'s clock state $z_{i}(t)$.}%
\label{z_dynamic}%
\end{figure}

\textbf{State Space: }In addition to $\mathcal{A}(t)$ and $\mathcal{P}(t)$
describing the state of the RSS, we define the states associated with each
vehicle and passenger as follows. Let $x_{j}(t)\in\mathbb{G}$ be the position
of vehicle $j$ at time $t$ and let $N_{j}(t)\in\{0,1,...,C_{j}\}$ be the
number of passengers in vehicle $j$ at time $t$, where $C_{j}$ is the capacity
of vehicle $j$. The state of passenger $i$ is denoted by $s_{i}(t)$ where
$s_{i}(t)=0$ if passenger $i$ is waiting to be picked up and $s_{i}%
(t)=j\in\mathcal{A}(t)$, where $j>0$, when the passenger is in vehicle $j$
after being picked up. Finally, we associate with passenger $i$ a
left-continuous clock value $z_{i}(t)\in$ $\mathbb{R}$ whose dynamics are
defined as follows: when the passenger joins the system and is added to
$\mathcal{P}(t)$, the initial value of $z_{i}(t)$ is $0$ and we set $\dot
{z}_{i}(t)=1$, as illustrated in Fig.\ref{z_dynamic} where the passenger
service request time is $\varphi_{i}$. Thus, $z_{i}(t)$ may be used to measure
the waiting time of passenger $i$. When $i$ is picked up by some vehicle $j$
at time $\rho_{i,j}$ (see Fig.\ref{z_dynamic}), $z_{i}(t)$ is reset to zero
and thereafter measures the traveling time until the passenger's destination
is reached at time $\sigma_{i,j}$. In summary, the state of the RSS is
$\mathbf{X}(t)=\{\mathcal{A}(t),x_{1}(t),\dots,x_{A}(t),N_{1}(t),\dots
,N_{A}(t),\mathcal{P}(t),s_{1}(t),\dots,s_{P}(t),\newline z_{1}(t),\dots
,z_{P}(t)\}$.

\textbf{Events: } All state transitions in the RSS are event-driven with the
exception of the passenger clock states $z_{i}(t)$, $i\in\mathcal{P}(t)$, in
which case it is the reset conditions (see Fig.\ref{z_dynamic}) that are
event-driven. As we will see, all control actions (to be defined) affecting
the state $\mathbf{X}(t)$ are taken only when an event takes place. Therefore,
regarding a vehicle location $x_{j}(t)$, $j\in\mathcal{A}(t)$, for control
purposes we are interested in its value only when events occur, even though we
assume that $x_{j}(t)$ is available to the RSS for all $t$ based on an
underlying localization system.

We define next the set $E$ of all events whose occurrence causes a state
transition. We set $E=E_{U}\cup E_{C}$ to differentiate between uncontrollable
events contained in $E_{U}$ and controllable events contained in $E_{C}$.
There are six possible event types, defined as follows:

(1) $\alpha_{i}\in E_{U}$: a service request is issued by passenger $i$.

(2) $\beta_{j}\in E_{U}$: vehicle $j$ joins the RSS.

(3) $\gamma_{j}\in E_{U}$: vehicle $j$ leaves the RSS.

(4) $\pi_{i,j}\in E_{C}$: vehicle $j$ picks up passenger $i$ (at $o_{i}%
\in\mathbb{G}$).

(5) $\delta_{i,j}\in E_{C}$: vehicle $j$ drops off passenger $i$ (at $r_{i}%
\in\mathbb{G}$).

(6) $\zeta_{m,j}\in E_{C}$: vehicle $j$ arrives at intersection (node)
$m\in\mathcal{N}$.

Note that events $\alpha_{i}$, $\beta_{j}$ are uncontrollable exogenous
events. Event $\gamma_{j}$ is also uncontrollable, however it may not occur
unless the \textquotedblleft guard condition\textquotedblright\ $N_{j}(t)=0$
is satisfied, that is, the number of passengers in vehicle $j$ must be zero
when it leaves the system. On the other hand, the remaining three events are
controllable. First, $\pi_{i,j}$ depends on the control policy (to be defined)
through which a vehicle is assigned to a passenger and is feasible only when
$s_{i}(t)=0$ and $N_{j}(t)<C_{j}$. Second, $\delta_{i,j}$ is feasible only
when $s_{i}(t)=j\in\mathcal{A}(t)$. Finally, $\zeta_{m,j}$ depends on the
policy (to be defined) and occurs when the route taken by vehicle $j$ involves
intersection $m\in\mathcal{N}$.

\textbf{State Dynamics: }The events defined above determine the state dynamics
as follows.

(1) Event $\alpha_{i}$ adds an element to the passenger set $\mathcal{P}(t)$
and increases its cardinality, i.e., $P(t^{+})=P(t)+1$ where $t$ is the
occurrence time of this event. In addition, it initializes the passenger state
and associated clock:%
\begin{equation}
s_{i}(t^{+})=0,\text{ \ \ }\dot{z}_{i}(t^{+})=1\text{ with }z_{i}(t)=0
\label{alpha_dynamics}%
\end{equation}
and generates the origin information of this passenger $o_{i}\in\mathbb{G}$.

(2) Event $\beta_{j}$ adds an element to the vehicle set $\mathcal{A}(t)$ and
increases its cardinality, i.e., $A(t^{+})=A(t)+1$. It also initializes
$x_{j}(t)$ to the location of vehicle $j$ at time $t$.

(3) Event $\gamma_{j}$ removes vehicle $j$ from $\mathcal{A}(t)$ and decreases
its cardinality, i.e., $A(t^{+})=A(t)-1$.

(4) Event $\pi_{i,j}$ occurs when $x_{j}(t)=o_{i}$ and it generates the
destination information of this passenger $r_{i}\in\mathbb{G}$. This event
affects the states of both vehicle $j$ and passenger $i$:%
\[
N_{j}(t^{+})=N_{j}(t)+1,\text{ \ \ }s_{i}(t^{+})=j
\]
and, since the passenger was just picked up, the associated clock is reset to
$0$ and starts measuring traveling time towards the destination $r_{i}$:%
\begin{equation}
z_{i}(t^{+})=0,\text{ \ }\dot{z}_{i}(t^{+})=1 \label{pi _dynamics}%
\end{equation}

(5) Event $\delta_{i,j}$ occurs when $x_{j}(t)=r_{i}$ and it causes a removal
of passenger $i$ from $\mathcal{P}(t)$ and decreases its cardinality, i.e.,
$P(t^{+})=P(t)-1$. In addition, it affects the state of vehicle $j$:%
\[
N_{j}(t^{+})=N_{j}(t)-1
\]
\qquad

(6) Event $\zeta_{m,j}$ occurs when $x_{j}(t)=\nu_{m}$. This event triggers a
potential change in the control associated with vehicle $j$ as described next.

\textbf{Control: } The control we exert is denoted by $u_{j}(t)\in\mathbb{G}$
and sets the destination of vehicle $j$ in the RSS. We note that the
destination $u_{j}(t)$ may change while vehicle $j$ is en route to it based on
new information received as various events may take place. The control is
initialized when event $\beta_{j}$ occurs at some point $x_{j}(t)$ by setting
$u_{j}(t)=\nu_{m}$ where $m\in\mathcal{N}$ is the intersection closest to
$x_{j}(t)$ in the direction vehicle $j$ is headed. Subsequently, the vector
$\mathbf{u}(t)=\{u_{1}(t),\dots,u_{A}(t)\}$ is updated according to a given
policy whenever an event from the set $E$ occurs (we assume that all events
are observable by the RSS controller). Our control policy is designed to
optimize the objective function described next.

\textbf{Objective Function: } Our objective is to minimize the combined
\emph{waiting} and \emph{traveling} times of passengers in the RSS over a
given finite time interval $[0,T]$. In order to incorporate all passengers who
have received service over $[0,T]$, we define the set
\[
\mathcal{P}_{T}=\cup_{t\in\lbrack0,T]}\mathcal{P}(t)
\]
to include all passengers $i\in\mathcal{P}(t)$ for any $t\in\lbrack0,T]$. In
simple terms, $\mathcal{P}_{T}$ is used to record all passengers who are
either currently active in the RSS at $t=T$ or were active and departed at
some time $t<T$ when the associated $\delta_{i,j}$ event occurred for some
$j\in\mathcal{A}(t)$.

We define $w_{i}$ to be the waiting time of passenger $i$ and note that,
according to (\ref{alpha_dynamics}), $w_{i}=z_{i}(t)$ where $t$ is the time
when event $\pi_{i,j}$ occurs. Similarly, letting $y_{i}$ be the total
traveling time of passenger $i$, according to (\ref{pi _dynamics}) we have
$y_{i}=z_{i}(t)$ where $t$ is the time when event $\delta_{i,j}$ occurs. We
then formulate the following problem, given an initial state $\mathbf{X}_{0}$
of the RSS:
\begin{equation}
\min_{\mathbf{u}(t)}E\left[  \sum_{i\in\mathcal{P}_{T}}[\mu_{w}w_{i}+\mu
_{y}y_{i}]\right]  \label{cost_func1}%
\end{equation}
where $\mu_{w},\mu_{y}$ are weight coefficients defined so that $\mu_{w}%
=\frac{\omega}{W_{\max}}$ and $\mu_{y}=\frac{1-\omega}{Y_{\max}}$, $\omega
\in\lbrack0,1]$, and $W_{\max}$ and $Y_{\max}$ are upper bounds of the waiting
and traveling time of passengers respectively. The values of $W_{\max}$ and
$Y_{\max}$ are selected based on user experience to capture the worst case
tolerated for waiting and traveling times respectively. This construction
ensures that $w_{i}$ and $y_{i}$ are properly normalized so that
(\ref{cost_func1}) is well-defined.

The expectation in (\ref{cost_func1}) is taken over all random event times in
the RSS defined in an appropriate underlying probability space. Clearly,
modeling the random event processes so as to analytically evaluate this
expectation is a difficult task. This motivates viewing the RSS as unfolding
over time and adopting a control policy based on observed actual events and on
estimated future events that affect the RSS state.

\begin{figure}[pt]
\centering
\includegraphics[scale=0.35]{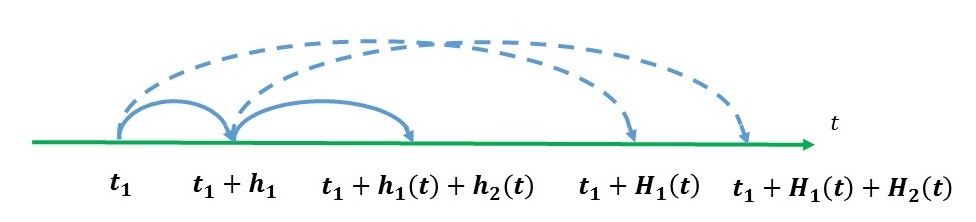}\caption{Event-Driven receding horizon control.}%
\label{RHC}%
\end{figure}

Assuming for the moment that the system is deterministic, let $t_{k}$ denote
the occurrence time of the $k$th event over $[0,T]$. A control action
$\mathbf{u}(t_{k})$ may be taken at $t_{k}$ and, for simplicity, is henceforth
denoted by $\mathbf{u}_{k}$. Along the same lines, we denote the state
$\mathbf{X}(t_{k})$ by $\mathbf{X}_{k}$. Letting $K_{T}$ be the number of
events observed over $[0,T]$, the optimal value of the objective function when
the initial state is $\mathbf{X}_{0}$ is given by
\[
J(\mathbf{X}_{0})=\min_{u_{0},\cdots u_{K_{T}}}\left[  \sum_{i\in
\mathcal{P}_{T}}[\mu_{w}w_{i}+\mu_{y}y_{i}]\right]
\]
We convert this into a maximization problem by considering $[-\mu_{w}w_{i}%
-\mu_{y}y_{i}]$ for each $i\in\mathcal{P}_{T}$. Moreover, observing that both
$w_{i}$ and $y_{i}$ are upper-bounded by $T$, we consider the non-negative
rewards $T-w_{i}$ and $T-w_{i}$ and rewrite the problem above as
\begin{equation}
J(\mathbf{X}_{0})=\max_{u_{0},\cdots u_{K_{T}}}\left[  \sum_{i\in
\mathcal{P}_{T}}[\mu_{w}(T-w_{i})+\mu_{y}(T-y_{i})]\right]
\label{Reward_Function}%
\end{equation}
Then, determining an optimal policy amounts to solving the following Dynamic
Programming (DP) equation \cite{bertsekas2005dynamic}:%
\[
J(\mathbf{X}_{k})=\max_{\mathbf{u}_{k}\in\mathbb{G}}[C(\mathbf{X}%
_{k},\mathbf{u}_{k})+J_{{k+1}}(\mathbf{X}_{k+1})],\text{ \ }k=0,1,\ldots,K_{T}%
\]
where $C(\mathbf{X}_{k},\mathbf{u}_{k})$ is the immediate reward at state
$\mathbf{X}_{k}$ when control $\mathbf{u}_{k}$ is applied and $J_{{k+1}%
}(\mathbf{X}_{k+1})$ is the future reward at the next state $\mathbf{X}_{k+1}%
$. Our ability to solve this equation is limited by the well-known
\textquotedblleft curse of dimensionality\textquotedblright%
\ \cite{bertsekas2005dynamic} even if our assumption that the RSS is fully
deterministic were to be valid. This further motivates adopting a
\emph{Receding Horizon Control} (RHC) approach as in similar problems
encountered in \cite{li2006cooperative} and \cite{khazaeni2016event}. This is
in the same spirit as Model Predictive Control (MPC) techniques
\cite{camacho2013model} with the added feature of exploiting the event-driven
nature of the control process. In particular, in the event-driven RHC
approach, a control action taken when the $k$th event is observed is selected
to maximize an immediate reward defined over a \emph{planning horizon} $H_{k}%
$, denoted by $C(\mathbf{X}_{k},\mathbf{u}_{k},H_{k})$, followed by an
estimated future reward $\hat{J}_{{k+1}}(\mathbf{X(}t_{k}+H_{k}))$ when the
state is $\mathbf{X(}t_{k}+H_{k})$. The optimal control action $\mathbf{u}%
_{k}^{\ast}$ is, therefore,%
\begin{equation}
\mathbf{u}_{k}^{\ast}=\arg\max_{\mathbf{u}_{k}\in\mathbb{G}}[C(\mathbf{X}%
_{k},\mathbf{u}_{k},H_{k})+\hat{J}_{{k+1}}(\mathbf{X(}t_{k}+H_{k}))]
\label{RHC_algorithm}%
\end{equation}
The control action $\mathbf{u}_{k}^{\ast}$ is subsequently executed only over
a generally shorter \emph{action horizon} $h_{k}\leq H_{k}$ so that
$t_{k+1}=t_{k}+h_{k}$ (see Fig.\ref{RHC}). The selection of $H_{k}$ and
$h_{k}$ will be discussed in the next section.

\section{Receding Horizon Control (RHC)}

In this section, we first review the basic RHC scheme as introduced in
\cite{li2006cooperative}, and a modified version in \cite{khazaeni2016event}
intended to overcome some of the original scheme's limitations. We refer to
the RHC in \cite{li2006cooperative} as \emph{RHC1} and the RHC in
\cite{khazaeni2016event} as \emph{RHC2}.

The basic RHC scheme in \cite{li2006cooperative} considers a set of
cooperating \textquotedblleft agents\textquotedblright\ and a set of
\textquotedblleft targets\textquotedblright\ in a Euclidean space. The purpose
of agents is to visit targets and collect a certain time-varying reward
associated with each target. The key steps of the scheme are as follows: (1)
Determine a planning horizon $H_{k}$ at the current time $t_{k}$. (2) Solve an
optimization problem to minimize an objective function defined over the time
interval $[t_{k},t_{k}+H_{k}]$. (3) Determine an action horizon $h_{k}$ and
execute the optimal solution over $[t_{k},t_{k}+h_{k}]$. (4) Set
$t_{k+1}=t_{k}+h_{k}$ and return to step (1).

Letting $\mathcal{A}(t)$ be the agent set and $\mathcal{P}(t)$ the target set,
we define $d_{i,j}(t)$ for any $i\in\mathcal{P}(t)$, $j\in\mathcal{A}(t)$ to
be the distance between target $i$ and agent $j$ at time $t$. In
\cite{li2006cooperative}, the planning horizon $H_{k}$ is defined as the
earliest time that any agent can visit any target in the system:
\begin{equation}
H_{k}=\min_{i\in\mathcal{P}(t),j\in\mathcal{A}(t)}\left\{  \dfrac
{d_{i,j}(t_{k})}{v}\right\}  \label{planninghorizon}%
\end{equation}
where $v$ is the fixed speed of agents. The action horizon $h_{k}$ is defined
to be the earliest time in $[t_{k},t_{k}+H_{k}]$ when an event in the system
occurs (e.g., a new target appears). In some cases, $h_{k}$ is alternatively
defined through $h_{k}=\epsilon H_{k}$ for some $\epsilon\in(0,1]$ so as to
ensure that $h_{k}\leq H_{k}$.

In order to formulate the optimization problem to be solved at every control
action point $t_{k}$, the concept of \emph{neighborhood} for a target is
defined in \cite{li2006cooperative} as follows. The $k$th nearest agent
neighbor to target $l$ is%
\[
\beta^{k}(l,t)=\underset{i\in\mathcal{A}(t),i\neq\beta_{l}^{1}(t),\dots
,i\neq\beta_{l}^{k-1}(t)}{\operatorname{arg}\,\operatorname{min}}%
\;{d_{l,i}(t)}%
\]
where $k=1,2,\ldots$, and the $b$-neighborhood of the target is given by the
set of the $b$ closest neighbors to it:
\begin{equation}
B_{l}^{b}(t)=\{\beta^{1}(l,t),\dots,\beta^{b}(l,t)\} \label{neighborset}%
\end{equation}
Based on \eqref{neighborset}, for any given $b\geq1$ the \emph{relative
distance} between agent $i$ and target $l$ is defined as
\begin{equation}
\bar{d}_{l,i}(t)=\left\{
\begin{array}
[c]{l}%
\dfrac{d_{l.i}(t)}{\sum_{q\in B_{l}^{b}(t)}d_{l,q}(t)}\\
1
\end{array}
\right.
\begin{array}
[c]{l}%
\text{if }i\in B_{l}^{b}(t)\\
\text{otherwise}%
\end{array}
\end{equation}
Then, the \emph{relative responsibility} function of agent $i$ for target $l$
is defined as:
\begin{equation}
p(\bar{d}_{l,i}(t))=\left\{
\begin{array}
[c]{l}%
1\\
\dfrac{1-\Gamma-\bar{d}_{l,i}}{1-2\Gamma}\\
0
\end{array}
\right.
\begin{array}
[c]{l}%
\text{if }\bar{d}_{l,i}\leq\Gamma\\
\text{if }\Gamma\leq\bar{d}_{l,i}\leq1-\Gamma\\
\text{otherwise}%
\end{array}
\label{RelativeResponsibility}%
\end{equation}
where $p(\bar{d}_{l,i}(t))$ can be viewed as the probability that agent $i$ is
the one to visit target $l$. In particular, when the relative distance is
small, then $i$ is committed to visit $l$, whereas if the relative distance is
large, then $i$ takes no responsibility for $l$. All other cases define a
\textquotedblleft cooperative region\textquotedblright\ where agent $i$ visits
$l$ with some probability dependent on the parameter $\Gamma$ which is
selected so that $\Gamma\in\lbrack0,\frac{1}{2})$ and reflects a desired level
of cooperation among agents; this cooperation level increases as $\Gamma$ decreases.

The use of $p(\bar{d}_{l,i}(t))$ allows the RHC to avoid early commitments of
agents to target visits, since changes in the system state may provide a
better opportunity for an agent to improve the overall system performance. A
typical example arises when agent $i$ is committed to target $l$ and a new
target, say $l^{\prime}$, appears which is in close proximity to $i$; in such
a case, it may be beneficial for $i$ to visit $l^{\prime}$ and let $l$ become
the responsibility of another agent that may be relatively close to $l$ and
uncommitted. This is possible if $p(\bar{d}_{l,i}(t))<1$. In what follows, we
will generalize the definition of distance $d_{i,j}(t)$ between target $i$ and
agent $j$ to the distance between any two points $x,y\in\mathbb{R}^{2}$
expressed as ${d(x,y)}$.

Using the relative responsibility function, the optimization problem solved by
the RHC at each control action point assigns an agent to a point which
minimizes a given objective function and which is not necessarily a target
point. Details of how this problem is set up and solved and the properties of
the \emph{RHC1} scheme may be found in \cite{li2006cooperative}.

\textbf{Limitations of }\emph{RHC1}\textbf{: }There are three main limitations
of the original RHC scheme:

\textit{(1) Agent trajectory instabilities:} A key benefit of \emph{RHC1} is
the fact that early commitments of agents to targets are avoided. As already
described above, if a new target appears in the system, an agent en route to a
different target may change its trajectory to visit the new one if this is
deemed beneficial to the cooperative system as a whole. This benefit, however,
is also a cause of potential instabilities when agents frequently modify their
trajectories, thus potentially wasting time. It is also possible that an agent
may oscillate between two targets and never visit either one. In
\cite{li2006cooperative}, necessary and sufficient conditions were provided
for some simple cases to quantify such instabilities, but these conditions may
not always be satisfied.

\textit{(2) Future cost estimation inaccuracies:} The effectiveness of
\emph{RHC1} rests on the accuracy of the future cost estimation term $\hat
{J}_{{k+1}}(\mathbf{X(}t_{k}+H_{k})$ in (\ref{RHC_algorithm}). In
\cite{li2006cooperative}, this future cost is estimated through its lower
bound, thus resulting in an overly \textquotedblleft
optimistic\textquotedblright\ outlook.

\textit{(3) Algorithm complexity:} In \cite{li2006cooperative}, the
optimization problem at each algorithm iteration involves the selection of
each agent's heading over $[0,2\pi]$. This is because the planning horizon
$H_{k}$ defines a set of feasible reachable points $F_{j}(t_{k},H_{k}%
)=\{w:d(w,x_{j}(t_{k})=vH_{k}\}$ which is a disk of radius $H_{k}/v$ (where
$v$ is each agent's speed) around the agent's position at time $t_{k}$. This
problem must be solved over all agents and incurs considerable computational
complexity: if $[0,2\pi]$ is discretized with discretization level $G$, then
the complexity of this algorithm at each iteration is $O(G^{A(t)})$.

The modified RCH scheme \emph{RHC2} in \cite{khazaeni2016event} was developed
to address these limitations. To deal with issues \emph{(1)} and \emph{(3)}
above, a set of \emph{active targets} $S_{j}(t_{k},H_{k})$ is defined for
agent $j$ at each iteration time $t_{k}$. Its purpose is to limit the feasible
reachable set $F_{j}(t_{k},H_{k})$ defined by all agent headings over
$[0,2\pi]$ so that it is reduced to a finite set of points. Let $x\in
F_{j}(t_{k},H_{k})$ be a reachable point and define a \emph{travel cost}
function $\eta_{i}(x,t)$ associated with every target $i\in\mathcal{P}(t)$
measuring the cost of traveling from a point $x$ at time $t$ to a target
$i\in\mathcal{P}(t)$. The active target set is defined in
\cite{khazaeni2016event} as
\begin{equation}
\begin{aligned} S_{j}(t_{k},H_{k})=\{l:l=\arg\min_{i\in\mathcal{P}(t)}\;\eta_{i}(x,t_{k}+H_{k})\\ \text{ for some }x\in F_{j}(t_{k},H_{k})\} \end{aligned} \label{ActiveTragets_RHC2}%
\end{equation}
Clearly, $S_{j}(t_{k},H_{k})\subseteq\mathcal{P}(t)$ is a finite set of
targets defined by the following property: an active target is closer to some
reachable point $x$ than any other target in the sense of minimizing the
metric $\eta_{i}(x,t_{k}+H_{k})$. Therefore, if there is some target
$l^{\prime}\notin S_{j}(t_{k},H_{k})$, then there is no incentive in
considering it as a candidate for agent $j$ to head towards. Restricting the
feasible headings of an agent to its active target set not only reduces the
complexity of optimally selecting a heading at $t_{k}$, but it also limits
oscillatory trajectory behavior, since by (\ref{planninghorizon}) there is
always an active target on the set $F_{j}(t_{k},H_{k})$ so that eventually all
targets are guaranteed to be visited.

Let $\mathbf{u}_{k}$ be the control applied at time $t_{k}$ under planning
horizon $H_{k}$. The $j$th component of $\mathbf{u}_{k}$ is the control
$u_{j}(t_{k})$ applied to agent $j$, where $u_{j}(t_{k})\in S_{j}(t_{k}%
,H_{k})$ as defined in (\ref{ActiveTragets_RHC2}). The estimated time for
agent $j$ to reach a target $u_{j}(t_{k})$ is denoted by $\hat{\tau}%
_{u,j}(\mathbf{u}_{k},t_{k},H_{k})$ where (for notational simplicity) we set
$u_{j}(t_{k})=u$. This time is given by%
\begin{equation}
\hat{\tau}_{u,j}(\mathbf{u}_{k},t_{k},H_{k})=t_{k}+H_{k}+\frac{1}{v}%
d(x_{j}(t_{k}),x_{u}),\text{ \ \ }u\in S_{j}(t_{k},H_{k}) \label{tauhat_1}%
\end{equation}
where $x_{u}$ is the location of target $u=u_{j}(t_{k})$.

To address issue \emph{(2)} regarding future cost estimation inaccuracies, a
new estimation framework is introduced in \cite{khazaeni2016event} by defining
a set of targets $\mathcal{T}_{k,j}\subseteq\mathcal{P}(t)-\{u\}$ that agent
$j$ would visit in the future, i.e., at $t>t_{k}+H_{k}$, as follows:
\begin{equation}
\mathcal{T}_{k,j}=\{l:p(\bar{d}_{l,j}(t_{k}))>p(\bar{d}_{l,q}(t_{k})),\text{
}\forall q\in\mathcal{A}(t)\} \label{targetset}%
\end{equation}
This set limits the targets considered by agent $j$ to those with a current
relative responsibility value in (\ref{RelativeResponsibility}) which exceeds
that of any other agent. The estimated time to reach a target $l\in
\mathcal{T}_{k,j}$ under control $\mathbf{u}_{k}$ and planning horizon $H_{k}$
is denoted by $\hat{\tau}_{l,j}(\mathbf{u}_{k},t_{k},H_{k})$. The first target
to be visited in $\mathcal{T}_{k,j}$, denoted by $l^{1}$, is the one with the
minimal travel cost from target $u\in S_{j}(t_{k},H_{k})$, i.e., $l^{1}%
=\arg\min_{l\in\mathcal{T}_{k,j}}\{\eta_{l}(x_{u},\hat{\tau}_{u,j}%
(\mathbf{u}_{k},t_{k},H_{k}))\}$. Then, all subsequent targets in
$\mathcal{T}_{k,j}-\{l^{1}\}$ are similarly ordered as $\{l^{2},l^{3}%
,\ldots\}$. Therefore, setting $\mathcal{T}_{k,j}^{n}=\mathcal{T}%
_{k,j}-\{l^{1},\ldots,l^{n-1}\}$, $n=2,\ldots,|\mathcal{T}_{k,j}|$, we have
\[
l^{n+1}=\arg\min_{l\in\mathcal{T}_{k,j}^{n}}\{\eta_{l}(x_{l^{n}},\hat{\tau
}_{l^{n},j}(\mathbf{u}_{k},t_{k},H_{k}))\},\text{ \ }n=1,\ldots,|\mathcal{T}%
_{k,j}|
\]
and
\begin{equation}
\hat{\tau}_{l^{n+1},j}(\mathbf{u}_{k},t_{k},H_{k})=\hat{\tau}_{l^{n}%
,j}(\mathbf{u}_{k},t_{k},H_{k})+\frac{1}{v}d(x_{l^{n}},x_{l^{n+1}})
\end{equation}

\textbf{Limitations of the }\emph{RHC2}\textbf{ with respect to a RSS}:

\textit{(1) Euclidean vs. Graph topology: }Both \emph{RHC1} and \emph{RHC2}
are based on an underlying Euclidean space topology. In a RSS, however, we are
interested in a graph-based topology which requires the adoption of a
different distance metric.

\textit{(2) Future cost estimation inaccuracies}: The travel cost metric
$\eta_{i}(x,t)$ used in \emph{RHC2} assumes that all future targets to be
visited at $t>t_{k}+H_{k}$ are independent of each other and that an agent can
visit any target. However, in a RSS, each agent $j$ has a capacity limit
$C_{j}$. This has two implications: $(i)$ If a vehicle is full, it must first
be assigned to a drop-off point before it can visit a new pickup point, and
$(ii)$ The number of future pickup points is limited by $C_{j}-N_{j}(t)$, the
residual capacity of vehicle $j$.

The fact that there are two types of \textquotedblleft
targets\textquotedblright\ in a RSS (pickup points and drop-off points), also
induces an interdependence in the rewards associated with target visits.
Whereas in \cite{khazaeni2016event} a reward is associated with each target
visit, in a RSS the rewards are $w_{i}$ and $y_{i}$ where $y_{i}$ can only be
collected after $w_{i}$. This necessitates a new definition of the set
$\mathcal{T}_{k,j}$ in (\ref{targetset}). For example, if $i\in\mathcal{T}%
_{k,j}$ and vehicle $j$ is full and must drop off a passenger at a remote
location, then using (\ref{targetset}) would cause vehicle $j$ to first go to
the drop-off location and then return to pick up $i$; however, there may be a
free vehicle $k$ in the vicinity of $j$'s current location which is obviously
a better choice to assign to passenger $i$.

\textit{(3) Agent trajectory instabilities}: \emph{RHC2} does not resolve the
possibility of agent trajectory instabilities. Moreover, the nature of such
instabilities is different due to the graph topology used in a RSS.

In view of this discussion, we will present in the next section a new RHC
scheme specifically designed for a RSS and addressing the issues identified
above. We will keep using the term \textquotedblleft target\textquotedblright%
\ to refer to points $o_{i}$ and $r_{i}$ for all $i\in\mathcal{P}(t)$.

\section{The New RHC Scheme}

We begin by introducing some variables used in the new RHC scheme as follows.

(1) $d(u,v)$ is defined as the \emph{Manhattan distance}
\cite{farris1972estimating} between two points $u,v\in\mathbb{G}$. This
measures the shortest path distance between two points on a directed graph
that includes points on an arc of this graph which belong to $\mathbb{G}%
\subset\mathbb{R}^{2}$.

(2) $\mathcal{R}_{i,j}(t)$ is the set of the $n$ closest pickup locations in
the sense of the Manhattan distance defined above, where $n=C_{j}-N_{j}(t)-1$
if $j$ picks up $i$ at $o_{i}$ at time $t$, and $n=C_{j}-N_{j}(t)+1$ if $j$
drops off $i$ at $r_{i}$ at time $t$. Clearly, the set may contain fewer than
$n$ elements if there are insufficient pickup locations in the RSS at time $t$.

(3) $\hat{\mathcal{R}}_{i,j}(t)$ is the set of $n$ drop-off locations for $j$,
where $n=N_{j}(t)+1$ if $j$ picks up $i$ at $o_{i}$, and $n=N_{j}(t)-1$ if $j$
drops off $i$ at $r_{i}$.

(4) $\varphi_{i}$ and $\rho_{{i,j}}$ denote the occurrence time of events
$\alpha_{i}$ (passenger $i$ joins the RSS) and $\pi_{i,j}$ (pickup of
passenger $i$ by vehicle $j$) respectively.

In the rest of this section we present the new RHC scheme which overcomes the
issues previously discussed through four modifications: $(i)$ We define the
\textit{travel value} of a passenger for each vehicle considering the distance
between vehicles and passengers, as well as the vehicle's residual capacity.
$(ii)$ Based on the new travel value and the graph topology of the map, we
introduce a new \textit{active target set} for each vehicle during
$[t_{k},t_{k}+H_{k})$. This allows us to reduce the feasible solution set of
the optimization problem (\ref{RHC_algorithm}) at each iteration. $(iii)$ We
develop an improved future reward estimation mechanism to better predict the
time that a passenger is served in the future. $(iv)$ To address the potential
instability problem, a method to restrain oscillations is introduced in the
optimization algorithm at each iteration.

Each of these modifications is described below, leading to the new RHC scheme.
We begin by defining the planning horizon $H_{k}$ at the $k$th control update
consistent with (\ref{planninghorizon}) as%

\begin{equation}
H_{k}=\min_{i\in\mathcal{P}(t_{k}),j\in\mathcal{A}(t_{k})}\left\{
\frac{d(x_{j}(t_{k}),c_{i})}{v_{j}(t_{k})}\right\}  \label{Horizon}%
\end{equation}
where
\begin{equation}
c_{i}=\left\{
\begin{array}
[c]{l}%
o_{i}\\
r_{i}\\
\end{array}
\right.
\begin{array}
[c]{l}%
\text{if }s_{i}(t)=0\text{ and }N_{j}(t_{k})<C_{j}\\
\text{if }s_{i}(t)=j\\
\end{array}
\label{ci_def}%
\end{equation}
and $v_{j}(t_{k})$ is the maximal speed of vehicle $j$ at time $t_{k}$,
assumed to be maintained over $[t_{k},t_{k}+H_{k}]$. Thus, $H_{k}$ is the
shortest Manhattan distance from any vehicle location to any target (either
$o_{i}$ or $r_{i}$) at time $t_{k}$. Note that $c_{i}$ is undefined if
$s_{i}(t)=0$ and $N_{j}(t_{k})=C_{j}$. Formally, to ensure consistency, we set
$d(x_{j}(t_{k}),c_{i})=\infty$ if $s_{i}(t)=0$ and $N_{j}(t_{k})=C_{j}$ since
$o_{i}$ is not a valid pickup point for $j$ in this case.

The action horizon $h_{k}\leq H_{k}$ is defined by the occurrence of the next
event in $E$, i.e., $h_{k}=\tau_{k+1}-t_{k}$ where $\tau_{k+1}$ is the time of
the next event to occur after $t_{k}$. If no such event occurs over
$[t_{k},t_{k}+H_{k}]$, we set $h_{k}=H_{k}$.

\subsection{Vehicle Travel Value Function}

Recall that in \emph{RHC2} a travel cost function $\eta_{i}(x,t)$ was defined
for any agent measuring the cost of traveling from a point $x$ at time $t$ to
a target $i\in\mathcal{P}(t)$. In our case, we define instead a \emph{travel
value} measuring the reward (rather than cost) associated with a vehicle $j$
when it considers any passenger $i\in\mathcal{P}(t)$. There are three cases to
consider depending on the state $s_{i}(t)$ for any $i\in\mathcal{P}(t)$ as follows:

\emph{Case 1:} If $s_{i}(t)=0$, then passenger $i$ is waiting to be picked up.
From a vehicle $j$'s point of view, there are two components to the value of
picking up this passenger at point $o_{i}$: $(i)$ The accumulated waiting time
$t-\varphi_{i}$ of passenger $i$; the larger this waiting time, the higher the
value of this passenger is. $(ii)$ The distance of $j$ from $o_{i}$; the
shorter the distance, the higher the value of this passenger is. To ensure
this value component is non-negative, we define $D$ to be the largest possible
travel time between any two points in the RSS (often referred to as the
diameter of the underlying graph) and consider $D-d(x_{j}(t),o_{i})$ as this
value component.

In order to properly normalize each component and ensure its associated value
is restricted to the interval $[0,1]$, we use the waiting time upper bound
$W_{\max}$ introduced in (\ref{cost_func1}) and the distance upper bound $D$
to define the total travel value function as
\begin{equation}
V_{i,j}(x_{j}(t),t)=(1-\mu)\cdot\dfrac{t-\varphi_{i}}{W_{\max}}+\mu\cdot
\frac{D-d(x_{j}(t),o_{i})}{D} \label{vaa}%
\end{equation}
where $\mu\in\lbrack0,1]$ is a weight coefficient depending on the relative
importance the RSS places on passenger satisfaction (measured by waiting time)
and vehicle distance traveled. In the latter case, a large value of
$d(x_{j}(t),o_{i})$ implies that vehicle $j$ wastes time either traveling
empty (if $N_{j}(t)=0$) or adding to the traveling time of passengers already
on board (if $N_{j}(t)>0$).

\emph{Case 2:} If $s_{i}(t)=j\in\mathcal{A}(t)$, then passenger $i$ is already
on board with destination $r_{i}$. From vehicle $j$'s point of view, there are
again two components to the value of delivering this passenger to point
$r_{i}$: $(i)$ The accumulated travel time $t-\rho_{i,j}$ of passenger $i$.
$(ii)$ The distance of $j$ from $r_{i}$. Similar to (\ref{vaa}), we define
\begin{equation}
V_{i,j}(x_{j}(t),t)=(1-\mu)\cdot\dfrac{t-\rho_{i,j}}{Y_{\max}}+\mu\cdot
\frac{D-d(x_{j}(t),r_{i})}{D} \label{vbb}%
\end{equation}
where $Y_{\max}$ is the travel time upper bound introduced in
(\ref{cost_func1}).

\emph{Case 3:} If $s_{i}(t)=k\neq j$, $k\in\mathcal{A}(t)$, then passenger $i$
is already on board some other vehicle $k\neq j$. Therefore, from vehicle
$j$'s point of view, the value of this passenger is $V_{i,j}(x_{j}(t),t)=0$.

\begin{figure}[pt]
\centering
\includegraphics[scale=0.4]{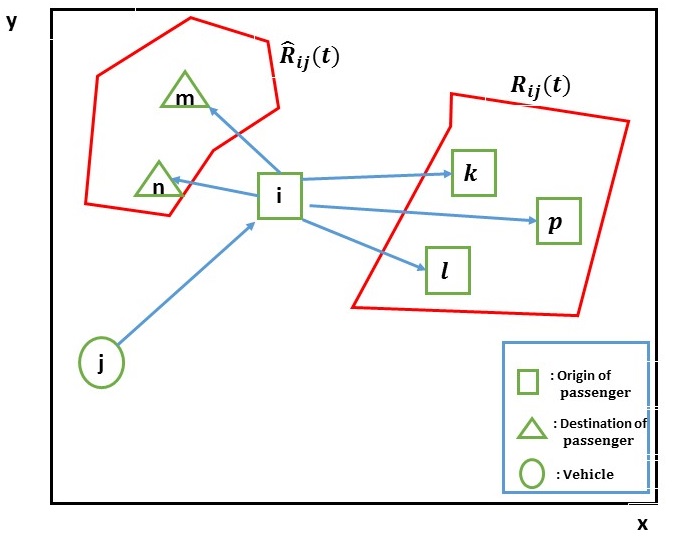} \caption{Travel value of
passenger $i$ evaluated by vehicle $j$ when $s_{i}(t)=0$.}%
\label{estimatefuturevalue}%
\end{figure}

We summarize the definition of the travel value function as follows:%
\begin{equation}
V_{i,j}(x_{j}(t),t)=\left\{
\begin{array}
[c]{cc}%
(1-\mu)\cdot\frac{t-\varphi_{i}}{W_{\max}}+\mu\cdot\frac{D-d(x_{j}(t),o_{i}%
)}{D} & \text{if }s_{i}(t)=0\\
(1-\mu)\cdot\frac{t-\rho_{i,j}}{Y_{\max}}+\mu\cdot\frac{D-d(x_{j}(t),r_{i}%
)}{D} & \text{if }s_{i}(t)=j\\
0 & \text{otherwise}%
\end{array}
\right.  \label{TravelValueFunction}%
\end{equation}
In addition to this \textquotedblleft immediate\textquotedblright\ value
associated with passenger $i$, there is a future value for vehicle $j$ to
consider depending on the sets $\mathcal{R}_{i,j}(t)$ and $\hat{\mathcal{R}%
}_{i,j}(t)$ defined earlier. In particular, if $s_{i}(t)=0$ and vehicle $j$
proceeds to the pickup location $o_{i}$, then the value associated with
$\mathcal{R}_{i,j}(t)$ is defined as
\[
V_{i,j}^{\mathcal{R}}(x_{j}(t),t)=\max_{n\in\mathcal{R}_{i,j}(t)}V_{n,j}%
(o_{i},t)
\]
which is the maximal travel value among all passengers in $\mathcal{R}%
_{i,j}(t)$ to be collected if vehicle $j$ selects $o_{i}$ as its destination
at time $t$. On the other hand, if $s_{i}(t)=j$ and vehicle $j$ proceeds to
the drop-off location $r_{i}$, then $V_{n,j}(o_{i},t)$ above is replaced by
$V_{n,j}(r_{i},t)$. Since the value of $s_{i}(t)$ is known to $j$, we will use
$c_{i}$ as defined in (\ref{ci_def}) and write%
\[
V_{i,j}^{\mathcal{R}}(x_{j}(t),t)=\max_{n\in\mathcal{R}_{i,j}(t)}V_{n,j}%
(c_{i},t)
\]
Similarly, the value of $\hat{\mathcal{R}}_{i,j}(t)$ is defined as
\[
V_{i,j}^{\hat{\mathcal{R}}}(x_{j}(t),t)=\max_{n\in\hat{\mathcal{R}}_{i,j}%
(t)}V_{n,j}(c_{i},t)
\]
We then define the total travel value associated with a vehicle $j$ when it
considers any passenger $i\in\mathcal{P}(t)$ as
\begin{equation}
\bar{V}_{i,j}(x_{j}(t),t)=V_{i,j}(x_{j}(t),t)+\max\{V_{i,j}^{\mathcal{R}%
}(x_{j}(t),t),V_{i,j}^{\hat{\mathcal{R}}}(x_{j}(t),t)\} \label{V_definition}%
\end{equation}
Figure \ref{estimatefuturevalue} shows an example of how $\bar{V}_{i,j}%
(x_{j}(t),t)$ is evaluated by vehicle $j$ in the case where $c_{i}=o_{i}$
(i.e., $s_{i}(t)=0$). In this case, $\mathcal{R}_{i,j}(t)=\{k,l,p\}$ and
$\hat{\mathcal{R}}_{i,j}(t)=\{m,n\}$.

\subsection{Active Target Sets}

The concept of an active target set was introduced in \cite{khazaeni2016event}%
. Clearly, this cannot be used in a RSS since the topology is no longer
Euclidean and the travel cost function $\eta_{i}(x,t)$ has been replaced by
the travel value function (\ref{V_definition}).

\begin{figure}[pt]
\centering
\includegraphics[height=5.5cm, width=8cm]{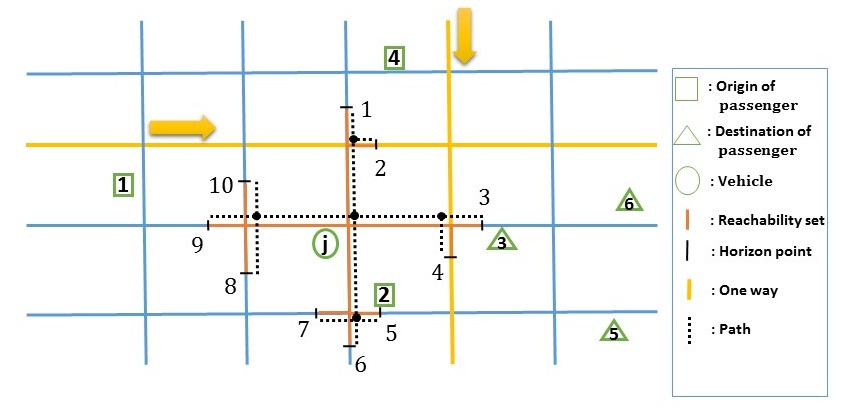} \caption{Example of the
reachability set of vehicle $j$.}%
\label{horizon_example}%
\end{figure}

We begin by defining the reachability (or feasible) set $F_{j}(t_{k},H_{k})$
for vehicle $j$ in the RSS topology specified by $\mathbb{G}\subset
\mathbb{R}^{2}$. This is now a finite set consisting of \emph{horizon points}
in $\mathbb{G}$ reachable through some path starting from $x_{j}(t_{k})$ and
assuming a fixed speed $v_{j}(t_{k})$ as defined in (\ref{Horizon}). This is
illustrated in Fig. \ref{horizon_example} where $F_{j}(t_{k},H_{k})$ consists
of 10 horizon points (one-way streets have been taken into account as directed
arcs in the underlying graph). Observe that $H_{k}$ in this example is defined
by $o_{2}$, the pickup location of passenger 2 (horizon point $5$) in
accordance with (\ref{Horizon}). Note that since the actual speed of the
vehicle may be lower than $v_{j}(t_{k})$, it is possible that no horizon point
is reached at time $t_{k}+h_{k}$ even if $h_{k}=H_{k}$. This simply implies
that a new planning horizon $H_{k+1}$ is evaluated at $t_{k}+H_{k}$ (which
might still be defined by $o_{2}$).We can now define the active target set of vehicle $j$ to consist
of any target (pickup or drop-off locations of passengers) which has the
largest travel value to $j$ for at least one horizon point $x\in F_{j}%
(t_{k},H_{k})$.

\textbf{Definition:} The set of \textit{Active Targets }of vehicle $j$ is
defined as\textit{ }
\begin{equation}
\begin{aligned} S_{j}(t_{k},H_{k})=\{l:l=\arg\max_{i\in\mathcal{P}(t)}\;\bar{V}_{i,j}(x,t_{k}+H_{k})\\ \text{ for some }x\in F_{j}(t_{k},H_{k})\} \end{aligned} \label{active_targets_RSS}%
\end{equation}
Observe that $S_{j}(t_{k},H_{k})\subseteq\mathcal{P}(t_{k})$ and may reduce
the number of passengers to consider as potential destinations assigned to $j$
when $S_{j}(t_{k},H_{k})\subset\mathcal{P}(t_{k})$ since
\[
u_{j}(t_{k})\in S_{j}(t_{k},H_{k})
\]
In the example of Fig. \ref{horizon_example}, $\mathcal{P}(t_{k})$ contains 6
passengers where $s_{1}(t_{k})=s_{2}(t_{k})=s_{4}(t_{k})=0$ and $s_{3}%
(t_{k})=s_{5}(t_{k})=s_{6}(t_{k})=j$. Thus, we can immediately see that
$P(t_{k}) =6<$ $\left\vert F_{j}(t_{k},H_{k})\right\vert =10$. Further,
observe that the drop-off points $r_{5}$ and $r_{6}$ are such that
$r_{5},r_{6}\notin S_{j}(t_{k},H_{k})$ since both points are farther away from
$x_{j}(t_{k})$ than $r_{3}$ and $o_{2}$ respectively. Therefore, the optimal
control selection to be considered at $t_{k}$ is reduced to $u_{j}(t_{k})\in
S_{j}(t_{k},H_{k})=\{o_{1},o_{2},r_{3},o_{4}\}$. In addition, if the capacity
$C_{j}$ happens to be such that $C_{j}=3$, then the only feasible control
would be $u_{j}(t_{k})=r_{3}$.



\subsection{Future Reward Estimation}

In order to solve the optimization problem (\ref{RHC_algorithm}) at each RHC
iteration time $t_{k}$, we need to estimate the time that a future target is
visited when $t>t_{k}+H_{k}$ so as to evaluate the term $\hat{J}_{{k+1}%
}(\mathbf{X(}t_{k}+H_{k}))$. Let us start by specifying the immediate reward
term $C(\mathbf{X}_{k},\mathbf{u}_{k},H_{k})$ in (\ref{RHC_algorithm}). In
view of (\ref{Reward_Function}), there are three cases: $(i)$ As a result of
$\mathbf{u}_{k}$, an event $\pi_{i,j}$ (where $s_{i}(t)=j$) occurs at time
$t_{k+1}$ with an associated reward $C(\mathbf{X}_{k},\mathbf{u}_{k}%
,H_{k})=\mu_{w}(T-w_{i})$ where $w_{i}=t_{k+1}-\varphi_{i}$, $(ii)$ As a
result of $\mathbf{u}_{k}$, an event $\delta_{i,j}$ occurs at time $t_{k+1}$
with an associated reward $C(\mathbf{X}_{k},\mathbf{u}_{k},H_{k})=\mu
_{y}(T-y_{i})$ where $y_{i}=t_{k+1}-\rho_{{i,j}}$, and $(iii)$ Any other event
results in no immediate reward. In summary, adopting the notation
$C(\mathbf{u}_{k},t_{k+1})$ for the immediate reward resulting from control
$\mathbf{u}_{k}$, we have
\begin{equation}
C(\mathbf{u}_{k},t_{k+1})=\left\{
\begin{array}
[c]{cc}%
\mu_{w}(T-w_{i}) & \text{if event }\pi_{i,j}\text{ occurs at }t_{k+1}\\
\mu_{y}(T-y_{i}) & \text{if event }\delta_{i,j}\text{ occurs at }t_{k+1}\\
0 & \text{otherwise}%
\end{array}
\right.  \label{Immediate_Reward}%
\end{equation}
In order to estimate future rewards at times $t>t_{k+1}$, recall that
$\mathcal{T}_{k,j}\subseteq\mathcal{P}(t)-\{u_{j}(t_{k})\}$ is a set of
targets that vehicle $j$ would visit in the future, after reaching
$u_{j}(t_{k})$. This set was defined in \cite{khazaeni2016event} through
(\ref{targetset}) and a new definition suitable for the RSS will be given
below. Then, for each target $n\in\mathcal{T}_{k,j}$ the associated reward is
$C(\mathbf{u}_{k},\hat{\tau}_{n,j})$ where $\hat{\tau}_{n,j}$ is the estimated
time that vehicle $j$ reaches target $n$. If $n=o_{i}$ for some passenger $i$,
then, from (\ref{Immediate_Reward}), $C(\mathbf{u}_{k},\hat{\tau}_{n,j}%
)=\mu_{w}(T-\hat{w}_{i})$ where $\hat{w}_{i}=\hat{\tau}_{n,j}-\varphi_{i}$,
whereas if $n=r_{i}$ for some passenger $i$, then $C(\mathbf{u}_{k},\hat{\tau
}_{n,j})=\mu_{y}(T-\hat{y}_{i})$ where $\hat{y}_{i}=\hat{\tau}_{n,j}%
-\rho_{{i,j}}$. Further, we include a \emph{discount factor} $\lambda_{n}%
(\hat{\tau}_{n,j})$ to account for the fact that the accuracy of our estimate
$\hat{\tau}_{n,j}$ is monotonically decreasing with time, hence $\lambda
_{n}(\hat{\tau}_{n,j})\in(0,1]$. Therefore, for each vehicle $j$ the
associated term for $\hat{J}_{{k+1}}(\mathbf{X(}t_{k}+H_{k}))$ is
\begin{equation}
\hat{J}_{j}(\mathbf{X}(t_{k}+H_{k}))=\sum_{n=1}^{\left\vert \mathcal{T}%
_{k,j}\right\vert }\lambda_{n}(\hat{\tau}_{n,j})C(u_{k,j},\hat{\tau}_{n,j})
\label{future_reward}%
\end{equation}
and%
\begin{equation}
\hat{J}(\mathbf{X}(t_{k}+H_{k}))=\sum_{j\in\mathcal{A}(t_{k})}\hat{J}%
_{j}(\mathbf{X}(t_{k}+H_{k})) \label{Total_future_reward}%
\end{equation}

We now need to derive estimates $\hat{\tau}_{n,j}$ for each $n\in
\mathcal{T}_{k,j}$. These estimates clearly depend on the order imposed on the
elements of $\mathcal{T}_{k,j}$, i.e., the expected order that vehicle $j$
follows in reaching the targets (after it reaches $u_{j}(t_{k})$) contained in
this set. As already explained under \emph{(2)} at the end of the last
section, this order depends on the passenger states and the residual capacity
of the vehicle. Suppose that the order is specified through $\theta_{n}^{j}$
defined as the $n$th target label in $\mathcal{T}_{k,j}$ (e.g., $\theta
_{1}^{j} = 4$ indicates that target $4$ is the first to be visited). Then,
(\ref{future_reward}) is rewritten as%
\begin{equation}
\hat{J}_{j}(\mathbf{X}(t_{k}+H_{k}))=\sum_{n=1}^{\left\vert \mathcal{T}%
_{k,j}\right\vert }\lambda_{\theta_{n}^{j}}(\hat{\tau}_{\theta_{n}^{j}%
,j})C(u_{k,j},\hat{\tau}_{\theta_{n}^{j},j}) \label{future_reward_ordered}%
\end{equation}

It now remains to $(i)$ define the set $\mathcal{T}_{k,j}$, suitably modified
from (\ref{targetset}) to apply to a RSS, so as to address the inaccuracy
limitation \emph{(2)} described at the end of the last section, and $(ii)$
Specify the ordering $\{\theta_{1}^{j},\ldots,\theta_{\left\vert
\mathcal{T}_{k,j}\right\vert }^{j}\}$ imposed on the elements of
$\mathcal{T}_{k,j}$.

We proceed by defining target subsets of $\mathcal{T}_{k,j}$ ordered in terms
of the priority of vehicle $j$ to visit these targets compared to other
vehicles. This is done using the relative responsibility function in
(\ref{RelativeResponsibility}) with the Manhattan distance used in evaluating
$\bar{d}_{l,i}(t)$. Thus, let $\mathcal{T}_{k,j}=\mathcal{T}_{k,j}^{1}%
\cup\dots\cup\mathcal{T}_{k,j}^{M}$ where $\mathcal{T}_{k,j}^{m}$ has the
$m$th highest priority among all subsets and $M\leq P(t)$ is the number of
subsets. When $m=1$, we have
\[
\mathcal{T}_{k,j}^{1}=\{l:p(\bar{d}_{l,j}(t_{k}))>p(\bar{d}_{l,q}%
(t_{k})),\text{ }\forall q\in\mathcal{A}(t),\text{ \ }\forall l\in
\mathcal{P}(t)\}
\]
which is the same as (\ref{targetset}): this is the passenger
\textquotedblleft responsibility set\textquotedblright\ of vehicle $j$ in the
sense that this vehicle has a higher responsibility value in
(\ref{RelativeResponsibility}) for each passenger in $\mathcal{T}_{k,j}^{1}$
than that of any other vehicle. Note that if $s_{l}(t_{k})=j$, then by default
we have $l\in T_{k,j}^{1}$ since the drop-off location $r_{i}$ is the
exclusive responsibility of vehicle $j$. For passengers with $s_{l}(t_{k})=0$,
they are included in $T_{k,j}^{1}$ as long as there is no other vehicle $q\neq
j$ with a higher relative responsibility for $l$ than that of $j$.

Next, let $\mathcal{A}_{l,m}(t)$ be a subset of vehicles defined as%
\[
\mathcal{A}_{l,m}(t_{k})=\{j:l\notin\mathcal{T}_{k,j}^{n},\text{ }n<m,\text{
\ }j\in\mathcal{A}(t_{k})\}
\]
This subset contains all vehicles which do not have target $l$ included in any
of their top $m-1$ priority subsets. We then define $\mathcal{T}_{k,j}^{m}$
when $m>1$ as follows:%
\begin{equation}
\begin{aligned} \mathcal{T}_{k,j}^{m}=\{l:p(\bar{d}_{l,j}(t_{k}))>p(\bar{d}_{l,q}(t_{k})),\text{ }\forall q\in\mathcal{A}_{l,m}(t_{k}),\\ \text{ }\forall l\notin\mathcal{T}_{k,j}^{n},n<m\text{\ }\} \end{aligned} \label{Tsets}%
\end{equation}
This set contains all targets for which $j$ has a higher relative
responsibility than any other vehicle and which have not been included in any
higher priority set $\mathcal{T}_{k,j}^{n},$ $n<m$. As an example, suppose
passenger $i$ is waiting to be picked up and belongs to $T_{k,j_{1}}^{1}$,
$T_{k,j_{2}}^{2}$ and $T_{k,j_{3}}^{3}$, where $j_{1}$ is the closest vehicle
to $i$. Suppose vehicle $j_{1}$ is full and needs to drop off a passenger
first whose destination is far away. Because vehicle $j_{2}$ has the $2$nd
highest priority, then $j_{2}$ may serve $i$ provided it has available seating
capacity. If $j_{2}$ cannot serve $i$, then vehicle $j_{3}$ with a lower
priority is the next to consider serving $i$. In this manner, we overcome the
limitation of (\ref{targetset}) where no agent capacity is taken into account.

The last step is to specify the ordering $\{\theta_{1}^{j},\ldots
,\theta_{\left\vert T_{k,j}^{m}\right\vert }^{j}\}$ imposed on each set
$T_{k,j}^{m},$ $j\in\mathcal{A}(t)$, $m=1,\ldots,M$. This is accomplished by
using the travel value function $\bar{V}_{i,j}(x_{j}(t),t)$ in
(\ref{V_definition}) as follows:
\begin{align}
\bar{V}_{\theta_{n+1}^{j},j}(c_{\theta_{n}^{j}},\hat{\tau}_{\theta_{n}^{j}%
,j})  &  \leq\bar{V}_{i,j}(c_{\theta_{n}^{j}},\hat{\tau}_{\theta_{n}^{j}%
,j})\label{v_last}\\
\text{for all }i  &  \in T_{k,j}^{m}-\{\theta_{1}^{j},\dots,\theta_{n}%
^{j}\}\nonumber
\end{align}
where we have used the definition of $c_{i}$ in (\ref{ci_def}). Setting
$u=u_{j}(t_{k})$, the estimated times are given by%
\begin{align}
\hat{\tau}_{\theta_{1}^{j},j}  &  =t_{k}+\frac{1}{v}d(x_{j}(t_{k}%
),x_{u})+\frac{1}{v}d(u,c_{\theta_{1}^{j},j})\label{tau1}\\
\hat{\tau}_{\theta_{n}^{j},j}  &  =\hat{\tau}_{\theta_{n-1}^{j},j}+\frac{1}%
{v}d(c_{\theta_{n-1}^{j},j},c_{\theta_{n}^{j},j}),\text{ \ }n>1 \label{taun}%
\end{align}
where $\hat{\tau}_{\theta_{1}^{j},j}$ is the estimated time of reaching the
target with the highest travel value beyond the one selected as $u_{j}(t_{k})$
among all targets in $T_{k,j}^{m}$ and $\hat{\tau}_{\theta_{n}^{j},j}$ for
$n>1$ is the estimated time of reaching the $n$th target in the order
established through (\ref{v_last}). Note that this approach takes into account
the state of vehicle $j$; in particular, if $N_{j}(t)=C_{j}$, then the
ordering of targets in $T_{k,j}^{m}$ is limited to those such that
$s_{i}(t_{k})=j$.

This completes the evaluation of the estimated future reward in
(\ref{Total_future_reward}) based on (\ref{Immediate_Reward}) and
(\ref{future_reward}), along with the ordering of future targets specified
through (\ref{v_last}).

\subsection{Preventing Vehicle Trajectory Instabilities}

Our final concern is the issue of instabilities discussed under \emph{(3)} at
the end of the last section. This problem arises when a new passenger joins
the system and introduces a new target for one or more vehicles in its
vicinity which may have higher travel value in the sense of
(\ref{V_definition}) than current ones. As a result, a vehicle may switch its
current destination $u_{j}(t_{k})$ and this process may repeat itself with
additional future new passengers. In order to avoid frequent such switches, we
introduce a threshold parameter denoted by $\Theta$ and react to any event
$\alpha_{i}$ (a service request issued by a new passenger $i$) that occurs at
time $t_{k}$ as follows:%
\begin{equation}
u_{j}(t_{k})=\left\{
\begin{array}
[c]{cc}%
o_{i} &
\begin{array}
[c]{c}%
\text{if }\bar{V}_{i,j}(x_{j}(t_{k}),o_{i})-\bar{V}_{u,j}(x_{j}(t_{k}%
),x_{u})>\Theta\text{,}\\
N_{j}(t)<C_{j}\text{, }j=1,\ldots,A(t_{k})
\end{array}
\\
u & \text{otherwise}%
\end{array}
\right.  \label{Threshold_Control}%
\end{equation}
where $u=u_{j}(t_{k-1})$ is the current destination of $j$. In simple terms,
the current control remains unaffected unless the new passenger provides an
incremental value relative to this control which exceeds a given threshold.
Since (\ref{Threshold_Control}) is applied to all vehicles in the current
vehicle set $\mathcal{A}(t)$, the vehicle with the largest incremental travel
value ends up with $o_{i}$ as its control as long as it exceeds $\Theta$. Note
that the new passenger may not be assigned to $j$ unless this vehicle has a
positive residual capacity.

\subsection{RHC optimization scheme}

The RHC scheme consists of a sequence of optimization problems solved at each
event time $t_{k}$, $k=1,2,\ldots$with each problem of the form%
\begin{equation}
\begin{aligned} \mathbf{u}_{k}^{\ast}=&\arg\max_{\substack{u_{k,j}\in S_{j}(t_{k},H_{k} )\\j\in\mathcal{A}(t_{k})}}[C(\mathbf{u}_{k},t_{k+1})\\ &+\sum_{j\in \mathcal{A}(t_{k})}\sum_{n=1}^{\left\vert T_{k,j}^{m}\right\vert }\lambda_{\theta_{n}^{j}}(\hat{\tau}_{\theta_{n}^{j},j})C(u_{k,j},\hat{\tau }_{\theta_{n}^{j},j})]\text{, \ }m=1,\ldots,M\label{RHCalgo}\end{aligned}
\end{equation}
where $S_{j}(t_{k},H_{k})$ is the active target of vehicle $j$ at time $t_{k}$
obtained through (\ref{active_targets_RSS}), $C(\mathbf{u}_{k},t_{k+1})$ is
given by (\ref{Immediate_Reward}), and $\hat{\tau}_{\theta_{n}^{j},j}$ is
evaluated through (\ref{tau1})-(\ref{taun}) with the ordering $\{\theta
_{1}^{j},\ldots,\theta_{\left\vert T_{k,j}^{m}\right\vert }^{j}\}$ given by
(\ref{v_last}) and the sets $\mathcal{T}_{k,j}^{m}$, $m=1,\ldots,M$, defined
through (\ref{Tsets}). Note that (\ref{RHCalgo}) must be augmented to include
(\ref{Threshold_Control}) when the event occurring at $t_{k}$ is of type
$\alpha_{i}$.

An algorithmic description of the RHC scheme is given in \textbf{Algorithm
$1$}

\begin{algorithm}
	\caption{RHC Algorithm.}\label{AG1}
	{ 1) Determine $H_k$ through (\ref{Horizon})\;}
	{ 2) Determine the active target set $S_j(t_k, H_k)$ through (\ref{active_targets_RSS}) for
		all $j\in \mathcal{A}(t)$\;}
	{ 3) Evaluate the estimated future reward through (\ref{tau1}) and (\ref{taun}) for all candidate optimal controls\;}
	
	{4) Determine the optimal control $\mathbf{u^*_k}$ in (\ref{RHCalgo})\;}
	
	{5) Execute $\mathbf{u^*_k}$ until an event occurs\;
	   	\If {a new passenger $i$ enters the system}{
			
			\For{each vehicle $j$ with $N_j(t)<C_j$}{	
				calculate $\bar{V}_{i,j}(x_{j}(t_{k}),o_{i})$\;
				\If{ $\bar{V}_{i,j}(x_{j}(t_{k}),o_{i})-\bar{V}_{u,j}(x_{j}(t_{k}%
					),x_{u})>\Theta$}{ we set $i$ as the new target\;
					break\;
				}
			}}}
			
		\end{algorithm}

\textbf{Complexity of Algorithm $1$}: The complexity of the original RHC in
\cite{li2006cooperative} was discussed in Section III. For the new RHC we have
developed, the optimal control for vehicle $j$ at any iteration is selected
from the finite set $S_{j}(t_{k},H_{k})$ defined by active targets. Thus, the
complexity is $O(\Omega^{A(t)})$ where $\Omega\leq P(t)$ (the number of
targets) is the maximum number of active targets. Observe that $\Omega$
decreases as targets are visited if new ones are not generated.

\bigskip

\section{Simulation Results}

We use the SUMO (Simulation of Urban Mobility) \cite{sumo} transportation
system simulator to evaluate our RHC for a RSS applied to two traffic networks
(in Ann Arbor, MI and in New York City, NY). Among other convenient features,
SUMO may be employed to simulate large-scale traffic networks and to use
traffic data and maps from other sources, such as OpenStreetMap and VISUM.
Vehicle speeds are set by the simulation and they include random factors like different road speed limits, turns, traffic lights, etc.
\begin{figure}[pt]
\centering
\includegraphics[height=6.5cm, width=8.5cm]{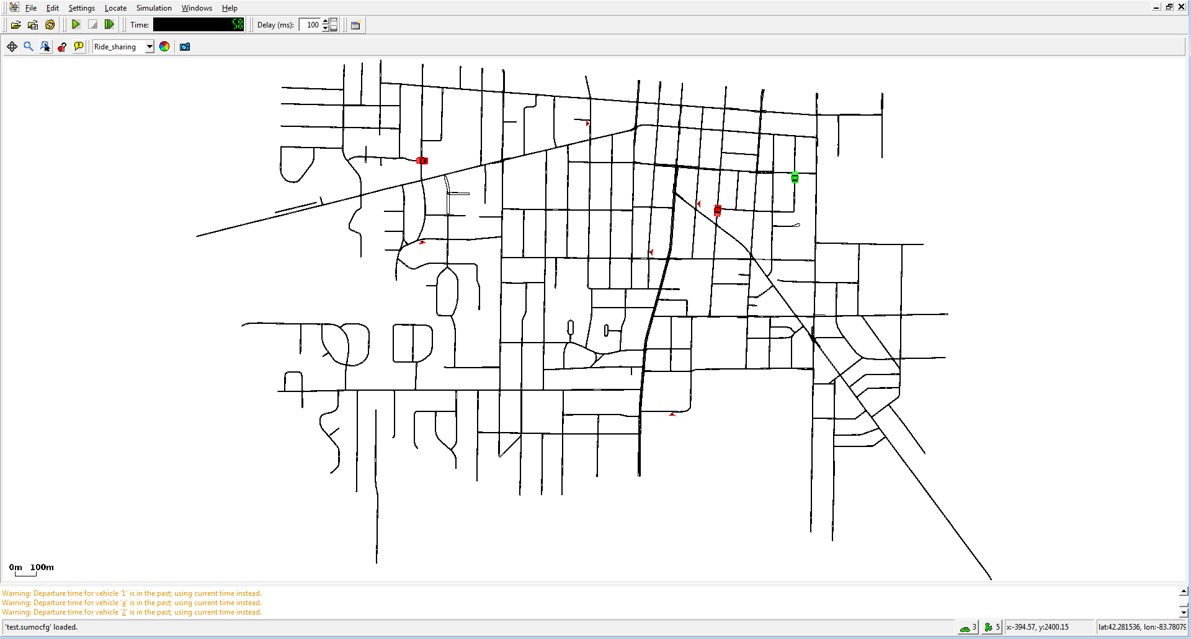} \caption{A RSS in the
Ann Arbor map.}%
\label{SUMO}%
\end{figure}

\subsection{RHC for a RSS in the Ann Arbor map}

A RSS for part of the Ann Arbor map is shown in Fig.\ref{SUMO}. Green colored
vehicles are idle while red colored ones contain passengers to be served. A
triangle along a road indicates a waiting passenger. We pre-load in SUMO a
fixed number of vehicles, while passengers request service at random points in
time as the simulation runs. Passenger arrivals are modeled as a Poisson
process with a rate of $3$ passengers/min. The remaining RSS system parameters
are selected as follows: $C_{j}=4$, $T=300$ min, $W_{max}=47$ min,
$Y_{max}=47$ min, $D=3000$ m and the threshold in (\ref{Threshold_Control}) is
set at $\Theta=0.3$. \begin{figure}[pt]
\centering
\includegraphics[height=12cm, width=8cm]{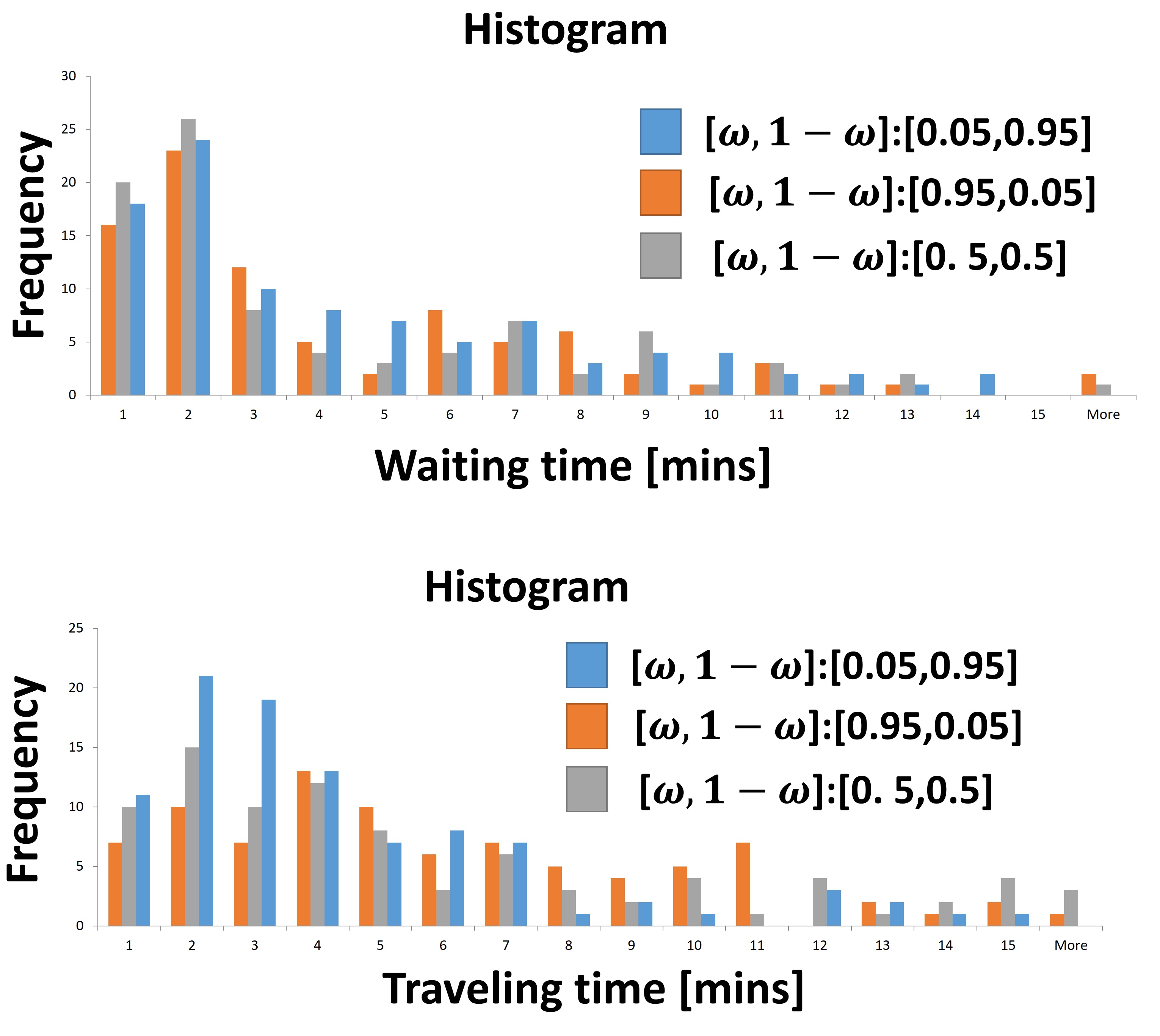} \caption{Waiting
and traveling time histograms under different weights $\omega$ for the Ann
Arbor RSS.}%
\label{AA_730_weight}%
\end{figure}

In Table \ref{tab:AA_730_weight}, the average waiting and traveling times
under RHC are shown for different weights $\omega$ in the Ann Arbor RSS. The
results are averaged over three independent simulation runs. In this example,
the number of pre-loaded vehicles is $7$ and simulations end after $30$
passengers are delivered to associated destinations (which is within T=300 min set above). In order to evaluate the performance of the RSS at steady state, we
allow a simulation to \textquotedblleft warm up\textquotedblright\ before
starting to measure the $30$ passengers served over the course of a simulation run.

The first column of Table \ref{tab:AA_730_weight} shows different values of
the weights $\omega$ as defined in \eqref{cost_func1} specifying the relative
importance assigned to passenger waiting and traveling respectively. As expected, emphasizing waiting results in
larger vehicle occupancy and longer average travel times. In Fig.
\ref{AA_730_weight} we provide the waiting and traveling time histograms for
all cases in Table \ref{tab:AA_730_weight}.

\begin{figure}[pt]
\centering
\includegraphics[height=10cm, width=8cm]{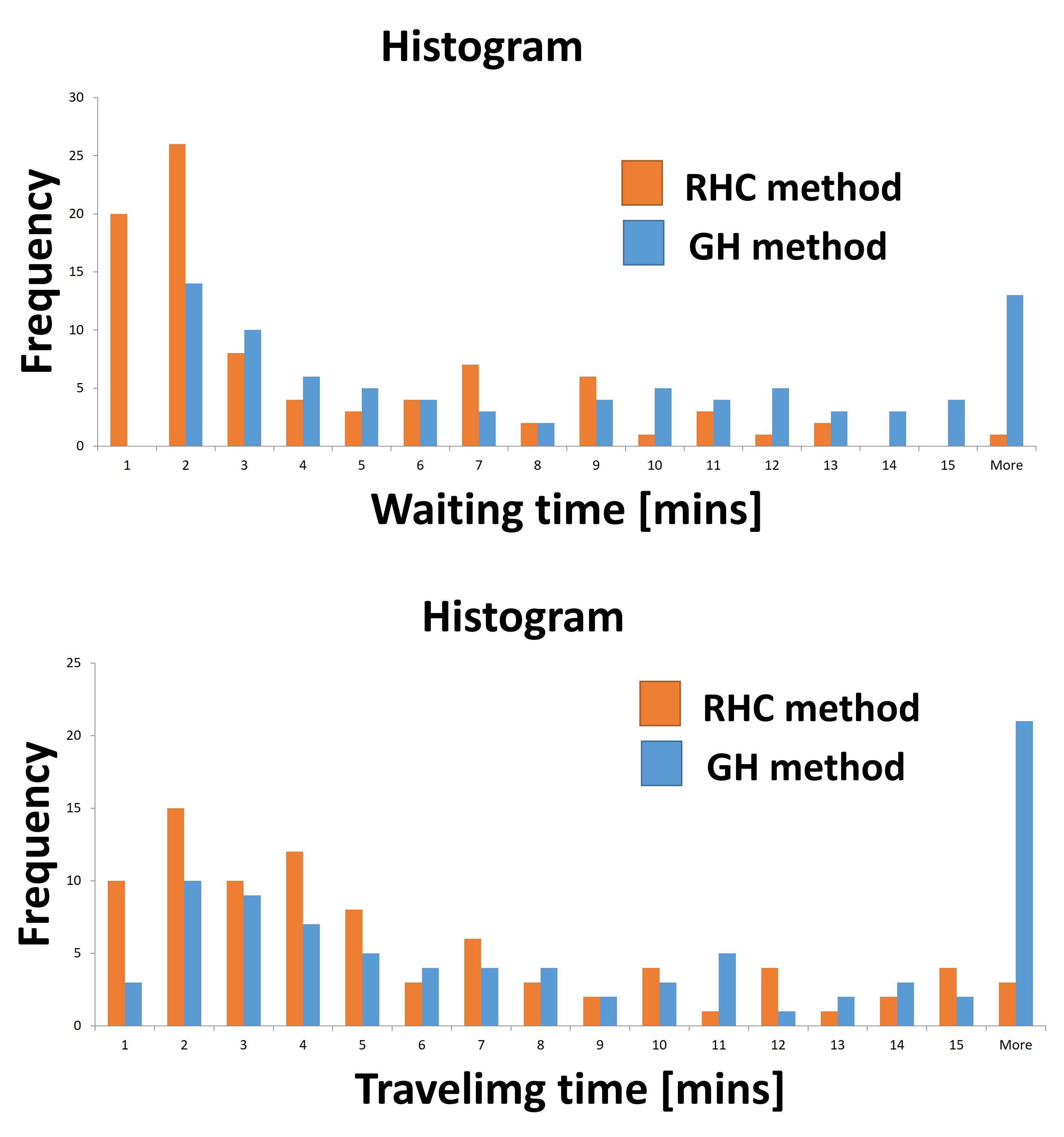}
\caption{Comparison of waiting and traveling time histograms under RHC and GH
for the Ann Arbor RSS ($\omega=0.5$).}%
\label{AA_730_Comparison}%
\end{figure}

In Table \ref{tab:AA_comparison}, we compare our RHC method with a greedy
heuristic (GH) algorithm (similar to \cite{agatz2011dynamic}) which operates as follows. When passenger $i$ joins
the RSS and generates the pickup point $o_{i}$, we evaluate the incremental
cost this point incurs to vehicle $j\in\mathcal{A}(t)$ when placed in every
possible position in this vehicle's current destination sequence, as long as
the capacity constraint $N_{j}(t)<C_{j}$ is never violated. The optimal
position is the one that minimizes this incremental cost. Once this is done
for all vehicles $j\in\mathcal{A}(t)$, we select the minimal incremental cost
incurred among all vehicles. Then, passenger $i$ is assigned to the associated
vehicle. As seen in Table \ref{tab:AA_comparison} with $\omega=0.5$, the RHC
algorithm achieves a substantially better weighted sum performance
 (approximately by a
factor of $2$) which are averaged over
three independent simulation runs. In Fig. \ref{AA_730_Comparison} we compare
the associated waiting and traveling time histograms showing in greater detail
the substantially better performance of RHC relative to GH.
Table \ref{tab:AA_comparison_vehicle_num} compares different vehicle numbers when the delivered passenger number is $30$ showing waiting and traveling times, vehicle occupancy and the objective in \eqref{cost_func1}. The larger the vehicle number, the better the performance can be achieved.

\begin{table}[ptb]
\caption{Average waiting and traveling times under RHC for different weights
$\omega$ in the Ann Arbor RSS}%
\label{tab:AA_730_weight}
\begin{center}
\resizebox{.45\textwidth}{.029\textheight}{
\begin{tabular}
[c]{l||c|c|c}\hline
$[\omega,1-\omega]$ & Waiting Time [mins] & Traveling Time [mins] & Vehicle
Occupancy\\\hline\hline
$[0.05,0.95]$ & 6.5 & 4.1 & 1.62\\\hline
$[0.5,0.5]$ & 6.0 & 5.2 & 2.64\\\hline
$[0.95,0.05]$ & 6.2 & 5.6 & 3.02\\\hline\hline
\end{tabular}}
\end{center}
\end{table}

\begin{figure}[pt]
\centering
\includegraphics[height=6.5cm, width=8cm]{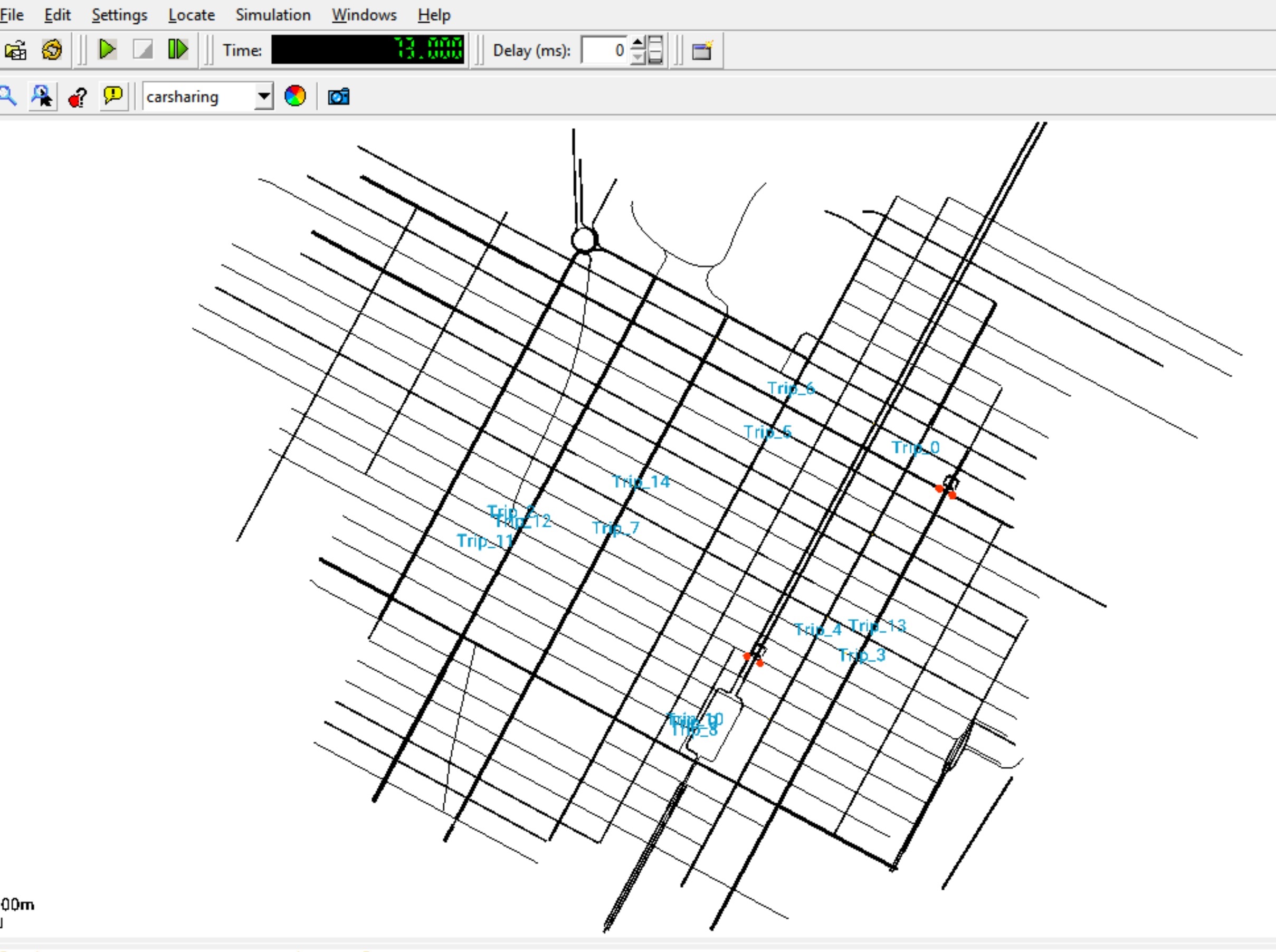} \caption{A RSS covering
an area of $10\times10$ blocks in New York City.}%
\label{RSS_NY}%
\end{figure}
\begin{table}[ptb]
\caption{Average waiting and traveling time [mins] comparisons for different
RSS control methods in the Ann Arbor RSS when $\omega=0.5$}%
\label{tab:AA_comparison}
\begin{center}%
		\resizebox{.45\textwidth}{.023\textheight}{
\begin{tabular}
[c]{l||c|c|c}\hline
Method & Waiting Time & Traveling Time & Weighted Sum in
\eqref{cost_func1}\\\hline\hline
RHC & 6.5 & 4.1 & 0.113\\\hline
GH & 9.6 & 9.7 & 0.205\\\hline
\end{tabular}}
\end{center}
\end{table}
\begin{table}[ptb]
	\caption{Average waiting and traveling time [mins] comparisons for different numbers of vehicles in the Ann Arbor RSS when $\omega=0.5$ under the RHC method}%
	\label{tab:AA_comparison_vehicle_num}
	\begin{center}%
		\resizebox{.45\textwidth}{.023\textheight}{
		\begin{tabular}
			[c]{c||c|c|c|c}\hline
			Vehicle Numbers & Waiting Time & Traveling Time & Vehicle Occupancy &Weighted Sum in
			\eqref{cost_func1}\\\hline\hline
				4 & 11.0 & 5.5 &2.93&0.176 \\\hline
			7 & 6.5 & 4.1 &2.64& 0.113\\\hline
		\end{tabular}}
	\end{center}
\end{table}

\subsection{RHC for a RSS in the New York City map}

A RSS covering an area of $10\times10$ blocks in New York City is shown in
Fig.\ref{RSS_NY}. In this case, we generate passenger arrivals based on actual
data from the NYC Taxi and Limousine Commission which provides exact timing of arrivals
and the associated origins and destinations.
We pre-loaded $8$ vehicles and run the simulations until $50$ passengers are
served based on actual data from a weekday of January, 2016 (the approximate passenger rate is $16$ passengers/min). All other RSS
settings are the same as before.

\begin{table}[ptb]
\caption{Average waiting and traveling times under RHC for different weights
$\omega$ in the New York City RSS with $8$ vehicles.}%
\label{tab:NYC_850_weight}
\begin{center}
\resizebox{.45\textwidth}{.029\textheight}{	
\begin{tabular}
[c]{l||c|c|c}\hline
$[\omega,1-\omega]$ & Waiting Time [mins] & Traveling Time [mins] & Vehicle
Occupancy\\\hline\hline
$[0.05,0.95]$ & 9.1 & 7.8 & 1.96\\\hline
$[0.5,0.5]$ & 11.9 & 9.0 & 2.59\\\hline
$[0.95,0.05]$ & 10.3 & 10.2 & 3.06\\\hline\hline
\end{tabular}}
\end{center}
\end{table}

\begin{figure}[pt]
\centering
\includegraphics[height=11cm, width=8cm]{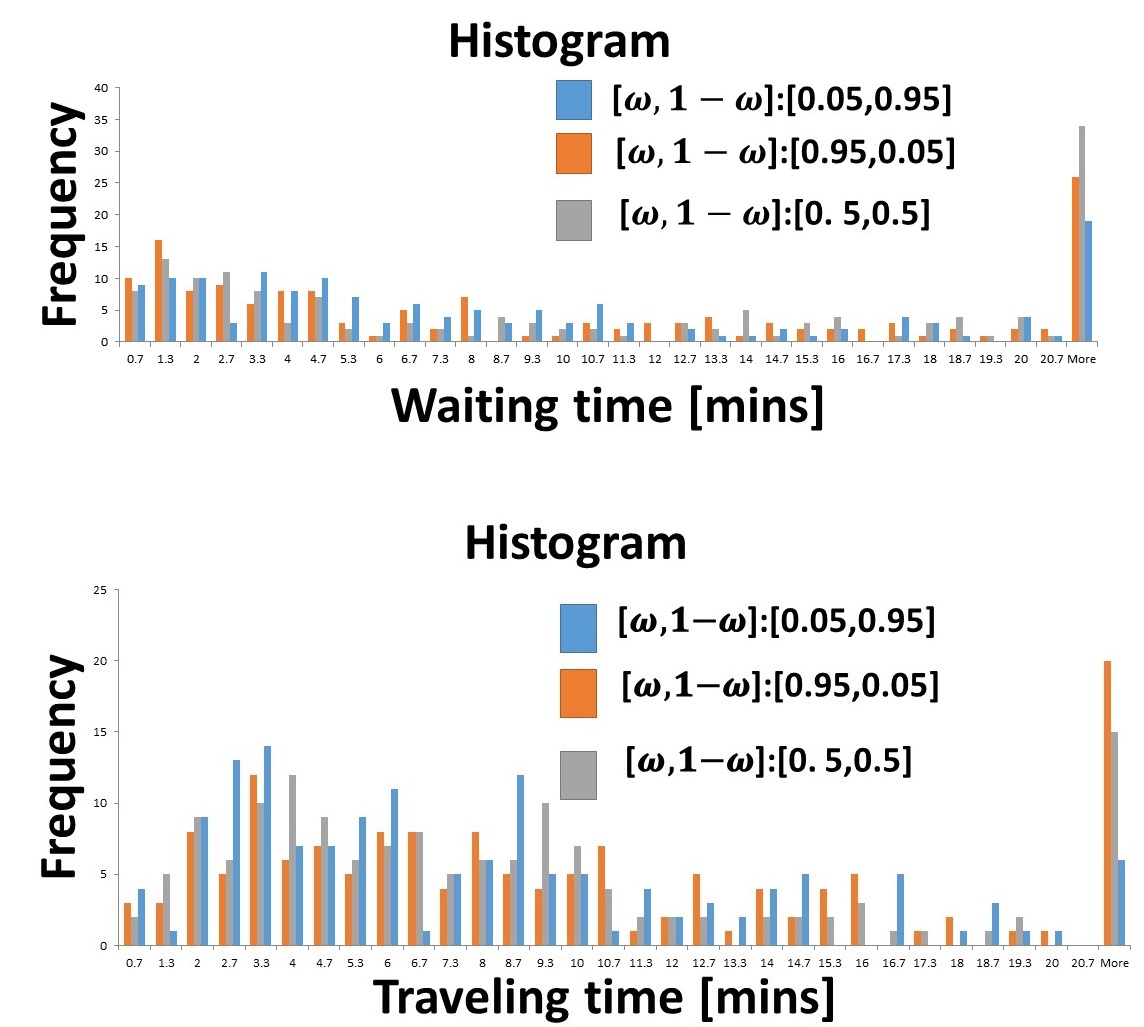} \caption{Waiting
and traveling time histograms under different weights $\omega$ for the New York
City RSS with $8$ vehicles.}%
\label{NYC_850_weight}%
\end{figure}\begin{table}[ptb]
\caption{Average waiting and traveling time [mins] comparisons for different
RSS control methods in the New York City RSS with $8$ vehicles and
$\omega=0.5$.}%
\label{tab:NYC_comparison}
\begin{center}%
		\resizebox{.45\textwidth}{.023\textheight}{
\begin{tabular}
[c]{l||c|c|c}\hline
Method & Waiting Time & Traveling Time & Weighted Sum in
\eqref{cost_func1}\\\hline\hline
RHC & 11.9 & 9.0 & 0.222\\\hline
GH &21.5  & 17.0 &0.410 \\\hline
\end{tabular}}
\end{center}
\end{table}\begin{figure}[pt]
\centering
\includegraphics[height=11cm, width=8cm]{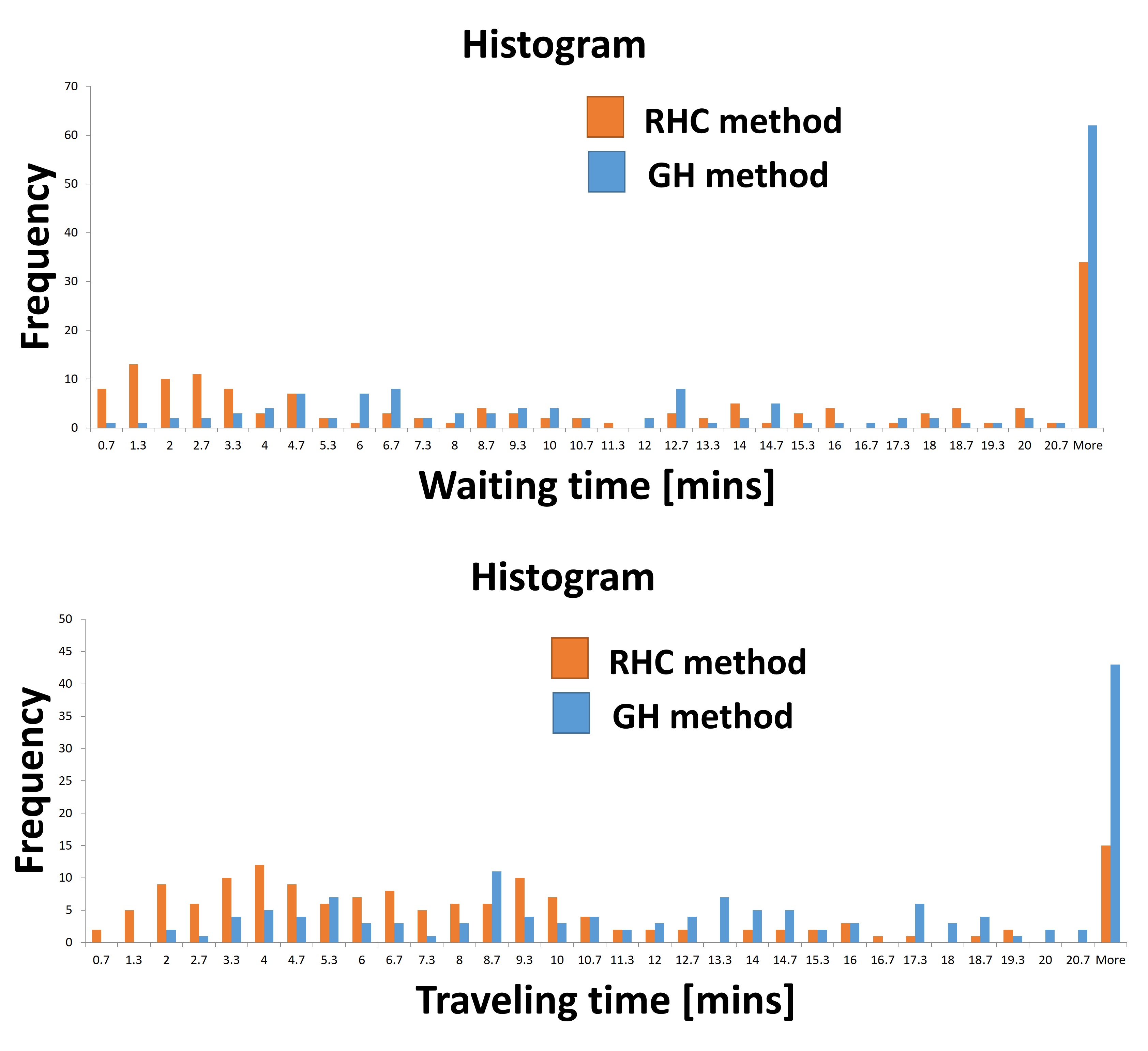}
\caption{Comparison of waiting and traveling time histograms under RHC and GH
in the New York City RSS with $8$ vehicles.}%
\label{NYC_850_Comparison}%
\end{figure}\begin{figure}[pt]
\centering
\includegraphics[height=11cm, width=8cm]{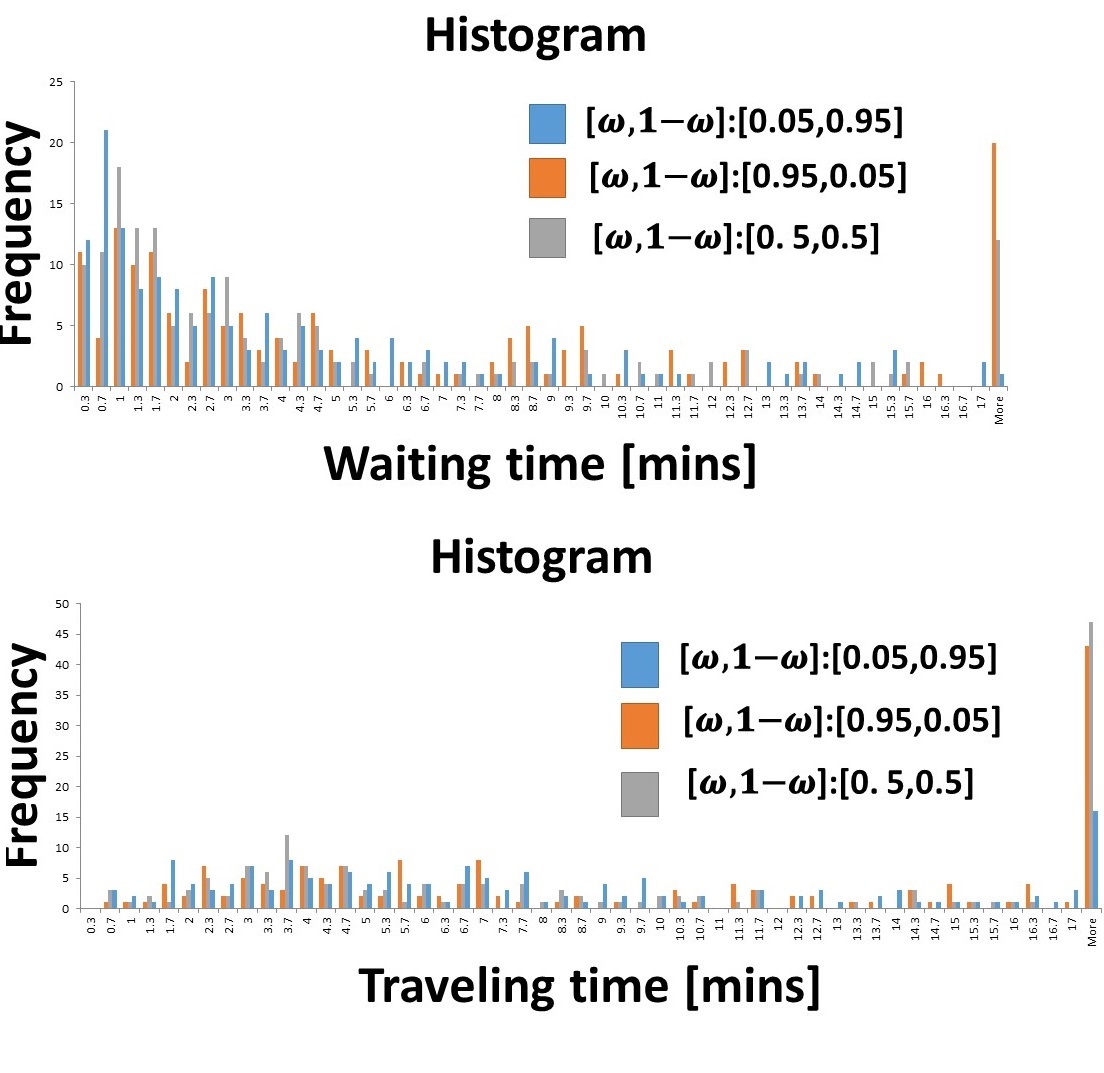}
\caption{Waiting and traveling time histograms under different weights
$\omega$ for the New York City RSS with $28$ vehicles.}%
\label{NYC_28160_weight}%
\end{figure}

\begin{table}[ptb]
\caption{Average waiting and traveling times under RHC for different weights
$\omega$ in the New York City RSS with $28$ vehicles.}%
\label{tab:NYC_28160_weight2}
\begin{center}
\resizebox{.45\textwidth}{.029\textheight}{
\begin{tabular}
[c]{l||c|c|c}\hline
$[\omega,1-\omega]$ & Waiting time [mins] & Traveling time [mins] & Vehicle
Occupancy\\\hline\hline
$[0.05,0.95]$ & 4.1 & 8.1 & 2.07\\\hline
$[0.5,0.5]$ & 5.2 & 12.4 & 2.79\\\hline
$[0.95,0.05]$ & 7.0 & 12.6 & 2.83\\\hline\hline
\end{tabular}}
\end{center}
\end{table}\begin{table}[ptb]
\caption{Average waiting and traveling time [mins] comparisons under RHC and
GH in the New York City RSS with $28$ vehicles and $\omega=0.5$. }%
\label{tab:NYC_comparison2}
\begin{center}%
		\resizebox{.45\textwidth}{.023\textheight}{
\begin{tabular}
[c]{l||c|c|c}\hline
Method & Waiting time & Traveling time & Weighted Sum in
\eqref{cost_func1}\\\hline\hline
RHC & 5.2 & 12.4 & 0.187\\\hline
GH &16.1 &16.6 &0.348\\\hline 
\end{tabular}}
\end{center}
\end{table}

\begin{figure}[pt]
\centering
\includegraphics[height=11cm, width=8cm]{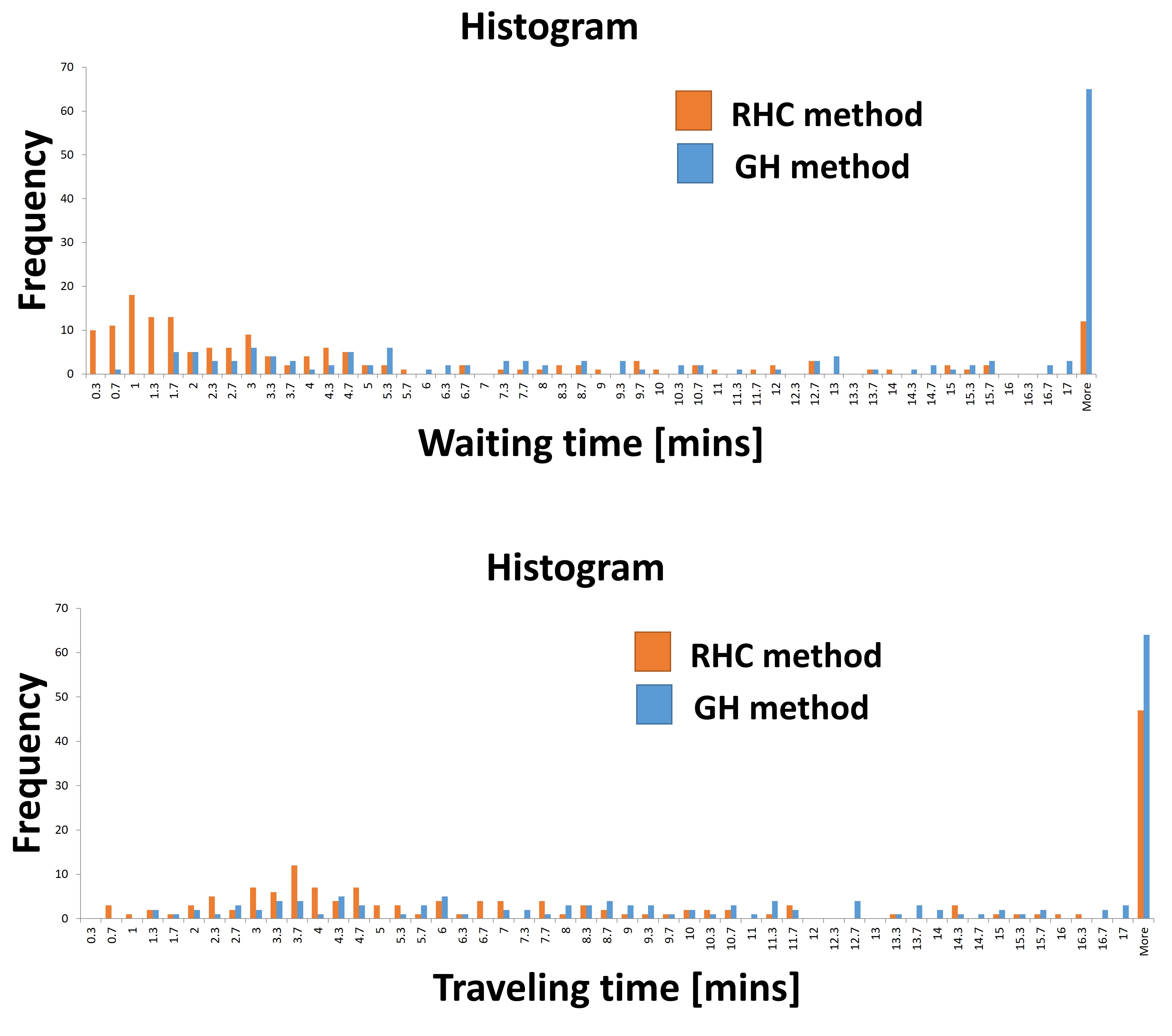}
\caption{Comparisons of waiting and traveling time histograms between the RHC
and GH methods in the New York City RSS when the vehicle number is $28$.}%
\label{NYC_28160_Comparison}%
\end{figure}

In Table \ref{tab:NYC_850_weight}, the average waiting and traveling times
under RHC are shown for different weights $\omega$ in the New York City RSS.
The results are averaged over three independent simulation runs. The first
column of Table \ref{tab:NYC_850_weight} shows different values of the weights
$\omega$ as defined in \eqref{cost_func1} specifying the relative importance
assigned to passenger waiting and traveling resepctively. As in the case of
the Ann Arbor RSS, emphasizing waiting results in larger vehicle occupancy
with longer average travel times. In Fig. \ref{NYC_850_weight} we provide the
waiting and traveling time histograms for all cases in Table
\ref{tab:NYC_850_weight}.

In Table \ref{tab:NYC_comparison}, we compare RHC with $\omega=0.5$ with the
aforementioned greedy heuristic algorithm GH in terms of the average waiting
and traveling times. We can see once again that the RHC algorithm achieves a
substantially better performance. 
In Fig.\ref{NYC_850_Comparison} we compare the associated waiting and traveling time
histograms for RHC relative to GH.

We have also tested a relatively long RSS operation based on actual passenger
data from a weekday of January 2016 which is the same as before for the
shorter time intervals. We pre-loaded $28$ vehicles and run simulations until
$160$ passengers are served. All other settings are the same as before.

Table \ref{tab:NYC_28160_weight2} shows the associated waiting and traveling
times under different weights with similar results as before. Figure
\ref{NYC_28160_weight} shows the associated waiting and traveling time
histograms for all cases in Table \ref{tab:NYC_28160_weight2}.

In Table \ref{tab:NYC_comparison2}, we compare RHC to the GH algorithm in
terms of the average waiting and traveling times with results consistent with
those of Table \ref{tab:NYC_comparison}. 

\begin{table}[ptb]
	\caption{Average waiting and traveling time [mins] comparisons for different numbers of vehicles in the New York City RSS when $\omega=0.5$ and the delivered passenger number is $160$ under the RHC method}%
	\label{tab:NYC_comparison_vehicle_num}
	\begin{center}%
		\resizebox{.47\textwidth}{.023\textheight}{
			\begin{tabular}
				[c]{c||c|c|c|c}\hline
				Vehicle Numbers & Waiting Time & Traveling Time & Vehicle Occupancy &Weighted Sum in
				\eqref{cost_func1}\\\hline\hline
				28 &5.2 & 12.4 & 2.79&0.187 \\\hline
				38 & 3.5 & 10.7 & 2.31&0.151\\\hline
		\end{tabular}}
	\end{center}
\end{table}

\begin{table}[ptb]
	\caption{Average real execution time for our RHC ALGO. when $\omega=0.5$}%
	\label{tab:RHC_assignment_time}
	\begin{center}%
				\resizebox{.45\textwidth}{.029\textheight}{
			\begin{tabular}
				[c]{c|c||c}\hline
				Vehicle Numbers &Passenger Numbers & Average Execution Time [sec] \\\hline\hline
								8 & 50 &3 \\\hline
				28 & 160 &17 \\\hline
				38 &160&19\\\hline
		\end{tabular}}
	\end{center}
\end{table}
Table \ref{tab:NYC_comparison_vehicle_num} compares different vehicle numbers when the delivered passenger number is $160$ showing waiting and traveling times, vehicle occupancy and the objective in \eqref{cost_func1} whose performance is consistent with that of Table \ref{tab:AA_comparison_vehicle_num}.

Table \ref{tab:RHC_assignment_time} shows real execution times for our RHC regarding different vehicle and passenger numbers. 
\begin{table}[ptb]
	\caption{Average waiting and traveling time [mins] comparisons under RHC and
		GH in the New York City RSS with $38$ vehicles and $\omega=0.5$. }%
	\label{tab:NYC_comparison3}
	\begin{center}
		\resizebox{.45\textwidth}{.023\textheight}{
			\begin{tabular}
				[c]{l||c|c|c}\hline
				Method & Waiting time & Traveling time & Weighted Sum in
				\eqref{cost_func1}\\\hline\hline
				RHC & 19.1 & 13.7 & 0.349\\\hline
				GH & 61.4& 19.0&0.855 \\\hline
		\end{tabular}}
	\end{center}
\end{table}

Finally, we tested a relatively longer RSS operation with $38$ vehicles based
on the same actual passenger data as before which generates $1000$ passengers over approximately $1.2$ 'real'
operation hours. Simulations will not end until $900$ passengers are
delivered. In Table \ref{tab:NYC_comparison3}, we compare RHC to the GH
algorithm in terms of the average waiting and traveling times with results
consistent with those of Table \ref{tab:NYC_comparison}.

\section{Conclusions and Future Work}
An event-driven RHC scheme is developed for a RSS where vehicles are shared to pick up ad drop off passengers so as to
minimize a weighted sum of passenger waiting and traveling times. The RSS is
modeled as a discrete event system whose event-driven nature significantly reduces the complexity of the vehicle assignment
problem, thus enabling its implementation in a real-time context. Simulation results adopting actual city maps and real taxi traffic data show the effectiveness of the RHC controller in terms of real-time implementation and performance relative to known greedy heuristics. In our ongoing work, an important problem we are considering is where to optimally position idle vehicles so that they are best used upon receiving future calls. Moreover, depending on real execution times of our RHC algorithm (see Table \ref{tab:RHC_assignment_time}), we will use this information as a rational measure for decomposing a map into regions such that within each region the RHC vehicle assignment response times remain manageable.

\bigskip
\bibliographystyle{IEEEtran}
\bibliography{CSS1206}

\begin{thebibliography}{10}
\providecommand{\url}[1]{#1}
\csname url@rmstyle\endcsname
\providecommand{\newblock}{\relax}
\providecommand{\bibinfo}[2]{#2}
\providecommand\BIBentrySTDinterwordspacing{\spaceskip=0pt\relax}
\providecommand\BIBentryALTinterwordstretchfactor{4}
\providecommand\BIBentryALTinterwordspacing{\spaceskip=\fontdimen2\font plus
\BIBentryALTinterwordstretchfactor\fontdimen3\font minus
  \fontdimen4\font\relax}
\providecommand\BIBforeignlanguage[2]{{%
\expandafter\ifx\csname l@#1\endcsname\relax
\typeout{** WARNING: IEEEtran.bst: No hyphenation pattern has been}%
\typeout{** loaded for the language `#1'. Using the pattern for}%
\typeout{** the default language instead.}%
\else
\language=\csname l@#1\endcsname
\fi
#2}}

\bibitem{schrank2011urban}
D.~Schrank, T.~Lomax, and B.~E. TTI’s, ``Urban mobility report. texas
  transportation institute, the texas a and m university system, 2007,'' 2011.

\bibitem{agatz2012optimization}
N.~Agatz, A.~Erera, M.~Savelsbergh, and X.~Wang, ``Optimization for dynamic
  ride-sharing: A review,'' \emph{European Journal of Operational Research},
  vol. 223, no.~2, pp. 295--303, 2012.

\bibitem{chen2017hierarchical}
X.~Chen, F.~Miao, G.~J. Pappas, and V.~Preciado, ``Hierarchical data-driven
  vehicle dispatch and ride-sharing,'' in \emph{Decision and Control (CDC),
  2017 IEEE 56th Annual Conference on}.\hskip 1em plus 0.5em minus 0.4em\relax
  IEEE, 2017, pp. 4458--4463.

\bibitem{agatz2011dynamic}
N.~A. Agatz, A.~L. Erera, M.~W. Savelsbergh, and X.~Wang, ``Dynamic
  ride-sharing: A simulation study in metro atlanta,'' \emph{Transportation
  Research Part B: Methodological}, vol.~45, no.~9, pp. 1450--1464, 2011.

\bibitem{santi2014quantifying}
P.~Santi, G.~Resta, M.~Szell, S.~Sobolevsky, S.~H. Strogatz, and C.~Ratti,
  ``Quantifying the benefits of vehicle pooling with shareability networks,''
  \emph{Proceedings of the National Academy of Sciences}, vol. 111, no.~37, pp.
  13\,290--13\,294, 2014.

\bibitem{berbeglia2010dynamic}
G.~Berbeglia, J.-F. Cordeau, and G.~Laporte, ``Dynamic pickup and delivery
  problems,'' \emph{European journal of operational research}, vol. 202, no.~1,
  pp. 8--15, 2010.

\bibitem{alonso2017demand}
J.~Alonso-Mora, S.~Samaranayake, A.~Wallar, E.~Frazzoli, and D.~Rus,
  ``On-demand high-capacity ride-sharing via dynamic trip-vehicle assignment,''
  \emph{Proceedings of the National Academy of Sciences}, vol. 114, no.~3, pp.
  462--467, 2017.

\bibitem{calafiore2017flow}
G.~C. Calafiore, C.~Novara, F.~Portigliotti, and A.~Rizzo, ``A flow
  optimization approach for the rebalancing of mobility on demand systems,'' in
  \emph{Decision and Control (CDC), 2017 IEEE 56th Annual Conference on}.\hskip
  1em plus 0.5em minus 0.4em\relax IEEE, 2017, pp. 5684--5689.

\bibitem{tsao2018stochastic}
M.~Tsao, R.~Iglesias, and M.~Pavone, ``Stochastic model predictive control for
  autonomous mobility on demand,'' \emph{arXiv preprint arXiv:1804.11074},
  2018.

\bibitem{salazar2018interaction}
M.~Salazar, F.~Rossi, M.~Schiffer, C.~H. Onder, and M.~Pavone, ``On the
  interaction between autonomous mobility-on-demand and public transportation
  systems,'' \emph{arXiv preprint arXiv:1804.11278}, 2018.

\bibitem{bertsekas2005dynamic}
D.~P. Bertsekas, D.~P. Bertsekas, D.~P. Bertsekas, and D.~P. Bertsekas,
  \emph{Dynamic programming and optimal control}.\hskip 1em plus 0.5em minus
  0.4em\relax Athena scientific Belmont, MA, 2005, vol.~1, no.~3.

\bibitem{camacho2013model}
E.~F. Camacho and C.~B. Alba, \emph{Model predictive control}.\hskip 1em plus
  0.5em minus 0.4em\relax Springer Science \& Business Media, 2013.

\bibitem{li2006cooperative}
W.~Li and C.~G. Cassandras, ``A cooperative receding horizon controller for
  multivehicle uncertain environments,'' \emph{IEEE Transactions on Automatic
  Control}, vol.~51, no.~2, pp. 242--257, 2006.

\bibitem{khazaeni2016event}
Y.~Khazaeni and C.~G. Cassandras, ``Event-driven cooperative receding horizon
  control for multi-agent systems in uncertain environments,'' \emph{IEEE
  Transactions on Control of Network Systems}, 2016.

\bibitem{farris1972estimating}
J.~S. Farris, ``Estimating phylogenetic trees from distance matrices,''
  \emph{The American Naturalist}, vol. 106, no. 951, pp. 645--668, 1972.

\bibitem{sumo}
\BIBentryALTinterwordspacing
G.~A.~C. (DLR). (2017) Simulation of urban mobility. [Online]. Available:
  \url{http://www.sumo.dlr.de/userdoc/Contact.html}
\BIBentrySTDinterwordspacing

\end{thebibliography}

\end{document}